# The JCMT SCUBA-2 Survey of the James Webb Space Telescope North Ecliptic Pole Time-Domain Field

Minhee Hyun[1,2], Myungshin Im[2], Ian R. Smail[3], William D. Cotton[4], Jack E. Birkin[3], Satoshi Kikuta[5], Hyunjin Shim[6], Christopher N. A. Willmer[7], James J. Condon[4], Rogier A. Windhorst[8], Seth H. Cohen[8], Rolf A. Jansen[8], Chun Ly[7], Yuichi Matsuda[9], Giovanni G. Fazio[10], A. M. Swinbank[3], and Haojing Yan[11]
[1] Korea Astronomy and Space Science Institute, 776 Daedeok-daero, Yuseong-gu, Daejeon 34055, Republic of Korea; minhee@kasi.re.kr
[2] SNU Astronomy Research Center, Astronomy Program, Department of Physics & Astronomy, Seoul National University, 1 Gwanak-ro, Gwanak-gu, Seoul 08826, Republic of Korea; myungshin.im@gmail.com
[3] Centre for Extragalactic Astronomy, Department of Physics, Durham University, South Road, Durham, DH1 3LE, UK
[4] National Radio Astronomy Observatory, 520 Edgemont Road, Charlottesville, VA 22903, USA
[5] Center for Computational Sciences, University of Tsukuba, Ten-nodai, 1-1-1 Tsukuba, Ibaraki 305-8577, Japan
[6] Department of Earth Science Education, Kyungpook National University, 80 Daehak-ro, Buk-gu, Daegu 41566, Republic of Korea
[7] Steward Observatory, University of Arizona, 933 North Cherry Avenue, Tucson, AZ 85721, USA
[8] School of Earth & Space Exploration, Arizona State University, Tempe, AZ 85287-1404, USA
[9] National Astronomical Observatory of Japan, Mitaka, Japan
[10] Center for Astrophysics, Harvard & Smithsonian, 60 Garden Street, MS-65, Cambridge, MA 02138-1516, USA
[11] Department of Physics and Astronomy, University of Missouri, Columbia, MO 65211, USA
*Received 2022 May 20; revised 2022 September 26; accepted 2022 September 26; published 2023 January 5*

## Abstract

The James Webb Space Telescope Time-Domain Field (JWST-TDF) is an ∼14′ diameter field near the North Ecliptic Pole that will be targeted by one of the JWST Guaranteed Time Observations programs. Here, we describe our James Clerk Maxwell Telescope SCUBA-2 850 $\mu$m imaging of the JWST-TDF and present the submillimeter source catalog and properties. We also present a catalog of radio sources from Karl J. Jansky Very Large Array 3 GHz observations of the field. These observations were obtained to aid JWST's study of dust-obscured galaxies that contribute significantly to cosmic star formation at high redshifts. Our deep 850 $\mu$m map covers the JWST-TDF at a noise level of $\sigma_{850\mu m} = 1.0$ mJy beam$^{-1}$, detecting 83/31 sources in the main/supplementary signal-to-noise ratio (S/N > 4 / S/N = 3.5–4) sample, respectively. The 3 GHz observations cover a 24′ diameter field with a 1$\sigma$ noise of 1 $\mu$Jy beam$^{-1}$ at a 0.″7 FWHM. We identified eighty-five 3 GHz counterparts to sixty-six 850 $\mu$m sources and then matched these with multiwavelength data from the optical to the mid-infrared wave bands. We performed spectral energy distribution fitting for 61 submillimeter galaxies (SMGs) matched with optical/near-infrared data, and found that SMGs at S/N > 4 have a median value of $z_{\mathrm{phot}} = 2.22 \pm 0.12$, star formation rates of $300 \pm 40\ M_\odot$ yr$^{-1}$ (Chabrier initial mass function), and typical cold dust masses of $5.9 \pm 0.7 \times 10^8\ M_\odot$, in line with bright SMGs from other surveys. The large cold dust masses indicate correspondingly large cool gas masses, which we suggest are a key factor necessary to drive the high star formation rates seen in this population.

*Unified Astronomy Thesaurus concepts:* Galaxy evolution (594); High-redshift galaxies (734); Galaxy formation (595); Submillimeter astronomy (1647); Galaxy counts (588); Ultraluminous infrared galaxies (1735)

*Supporting material:* extended figure, machine-readable tables

## 1. Introduction

Tracing the formation of galaxies and the associated star formation across cosmic time is a key topic for understanding the evolution of our universe. Most of the star formation activity over cosmic history is hidden by surrounding dust, making it often difficult to observe this activity through UV and optical studies (e.g., Madau & Dickinson 2014; Driver et al. 2016; Koushan et al. 2021). As such, it is important to study this hidden star formation in other wave bands to draw a complete picture of the evolution of galaxies, especially those in the early universe.

Dust around young stars efficiently absorbs their UV/optical radiation and this is reradiated in the mid-infrared (MIR) and far-infrared (FIR). Consequently, MIR and FIR surveys of galaxies have revealed a population of luminous infrared galaxies, whose contribution to the cosmic star formation rate (SFR) has been found to rise toward $z \sim 1.5$ (e.g., Goto et al. 2010, 2019; Magnelli et al. 2013; Kim et al. 2015). At higher redshifts, $z \gg 1.5$, the FIR emission from galaxies is redshifted into the submillimeter (submm) window. As a result, surveys in the submm are well placed to find strongly star-forming, dust-obscured galaxies at high redshifts, which are commonly termed submm galaxies (SMGs)—defined as galaxies that are bright in the submm, $S_{850\mu m} > 1$ mJy (Hodge & da Cunha 2020).

From their first discovery (Smail et al. 1997; Barger et al. 1998; Hughes et al. 1998; Eales et al. 1999), SMGs have been an important population for understanding the most obscured phases of galaxy formation and evolution at high redshifts. These galaxies are bolometrically luminous ($\sim 10^{12-13}\ L_\odot$; Barger et al. 1998) and massive ($\simeq 10^{11}\ M_\odot$; Hainline et al. 2011; Michałowski et al. 2012; Dudzevičiūtė et al. 2020) and contain a large amount of gas ($\sim 10^{11}\ M_\odot$; Greve et al. 2005; Tacconi et al. 2008; Carilli et al. 2010; Bothwell et al. 2013; Birkin et al. 2021). SMGs are considered the progenitors of elliptical or spheroidal galaxies in the local universe (Lilly et al. 1999; Miller et al. 2018) from the fact that they reside in massive dark matter halos with $M \sim 10^{12-13}\ M_\odot$ (Blain et al. 2004; Farrah et al. 2006; Magliocchetti et al. 2007; Wilkinson et al. 2017; Stach et al. 2021) and from their rapid formation (Lilly et al. 1999; Swinbank et al. 2006; Simpson et al. 2014). With the benefits of the strong negative $K$-correction in the submm, SMGs are relatively easy to identify even at very high redshifts ($z \sim 6$, e.g., Riechers et al. 2017; Gruppioni et al. 2020). They are now widely studied with various facilities and in numerous fields (Scott et al. 2002; Coppin et al. 2006; Weiß et al. 2009; Ikarashi et al. 2011; Geach et al. 2017; Simpson et al. 2019; Shim et al. 2022; see also Casey et al. 2014).

One of the key questions about SMGs is what physical mechanism triggers their star formation activity. We might expect that most SMGs arise from galaxy mergers based on studies in the local universe that show that the majority of ultraluminous infrared galaxies (ULIRGs) are merging systems (Sanders & Mirabel 1996; Lonsdale et al. 2006). Merging provides an effective mechanism to drive intense star formation and active galactic nucleus (AGN) activity (Barnes & Hernquist 1991; Mihos & Hernquist 1996; Hopkins et al. 2005; Springel et al. 2005; Hong et al. 2015; McAlpine et al. 2019). Many observational studies have found distorted morphologies for SMGs (Swinbank et al. 2004; Menendez-Delmestre et al. 2007; Tacconi et al. 2008; Engel et al. 2010; Chen et al. 2015) suggestive of mergers. Some semianalytic models also support the scenario that the vigorous star formation in SMGs is driven by galaxy merging (e.g., Chen et al. 2015). However, the very strong dust obscuration, even in the rest-frame near-infrared (NIR), makes it difficult to distinguish truly interacting systems from those where structured dust obscuration creates apparent asymmetric or disturbed morphologies. The stellar-mass morphologies of SMGs are one area where the recently launched James Webb Space Telescope (JWST) is expected to make significant contributions, through high-resolution MIR imaging that is much less sensitive to dust obscuration.

The JWST Time-Domain Field (JWST-TDF; Jansen & Windhorst 2018) is an $\sim 14'$ diameter region in JWST's northern continuous viewing zone near the North Ecliptic Pole (NEP). The JWST-TDF has a unique advantage as an extragalactic survey field in that it can be observed at any cadence with JWST and is also expected to be frequently visited due to spacecraft housekeeping activities. In addition, the field is free from sources brighter than $m_{AB} \sim 16$ and has low zodiacal foreground, as well as low Galactic extinction. The planned JWST observations (GTO 1176, PI: R. Windhorst) include eight-filter 0.8–5.0 $\mu$m imaging with the Near-Infrared Camera (NIRCam) and 1.75–2.23 $\mu$m grism spectroscopy and 2 $\mu$m direct imaging with the Near Infrared Imager and Slitless Spectrograph (NIRISS). These observations will detect sources down to $m_{AB} \simeq 29$ mag, and this sensitivity (and spatial resolution) makes the JWST-TDF ideal for exploring the morphologies of SMGs, especially those that are faint in the rest-frame optical/NIR.

To take advantage of the JWST-TDF for SMG studies, we have undertaken a submm survey covering the full JWST-TDF using the Submillimetre Common-user Bolometer Array 2 (SCUBA-2; Holland et al. 2013) of the James Clerk Maxwell Telescope (JCMT). We call this survey the SCUBA-2 JWST-TDF Survey (S2TDF) and to fully exploit its data we have also obtained sensitive, high-resolution 3 GHz observations of this region using the Karl J. Jansky Very Large Array (VLA). Together these two observational data sets support the wider multiwavelength studies by the JWST-TDF GTO team. In particular, we expect that our SCUBA-2 and VLA surveys of the JWST-TDF will provide critical information on the redshifted FIR and radio emission, and thus on the dust-obscured star formation rates and AGN activity, in galaxies to be detected by JWST and will be useful for the wider astronomical community for years to come. In this paper, we present the S2TDF, and the catalog of SCUBA-2 sources in this field, as well as details of the VLA 3 GHz survey of this region used to identify the counterparts to these SCUBA-2 sources. Section 2 presents the SCUBA-2 850 $\mu$m observations and the data reduction process (the corresponding description of the 3 GHz VLA observations and their reduction and cataloging is given in Appendix A). In Section 3, we describe the data analysis methods used to construct the catalog of 850 $\mu$m sources, such as jackknife simulations for deriving flux deboosting, completeness, the false detection rate, and positional accuracy. In Section 4, we discuss the matching of the SCUBA-2 sources with the 3 GHz radio and optical/NIR data to robustly identify the SMG counterparts of the SCUBA-2 sources, followed by the analysis of the spectral energy distributions (SEDs) of these counterparts using the available multiwavelength data. Section 5 presents the 850 $\mu$m number counts in the JWST-TDF and examines the properties of the radio- and optical/NIR-matched SMGs such as their redshifts, SFRs, and dust masses. In Section 6, we summarize our results. Throughout this paper, all magnitudes are given in the AB magnitude system (Oke & Gunn 1983) and we adopt cosmological parameters of $H_0 = 70$ km s$^{-1}$ Mpc$^{-1}$, $\Omega_\Lambda = 0.7$, and $\Omega_M = 0.3$. For the derivation of SMG properties with SED fitting, we use the Chabrier initial mass function (Chabrier 2003).

## 2. Observation and Data Reduction

In the following section we describe in detail the SCUBA-2 observations of the JWST-TDF. The corresponding description of the VLA 3 GHz observations is given in Appendix A.

### 2.1. SCUBA-2 Observations

The SCUBA-2 850 $\mu$m data were obtained from 2018 December to 2020 December (programs: M18BP026, M19BP031, and R20XP003) with the SCUBA-2 (Holland et al. 2013), the submm continuum bolometer array mounted on the JCMT. We used the PONG900 mapping mode, which generates a 15′ diameter circular map with nearly uniform sensitivity (Chapin et al. 2013). As a consequence, our data cover the full area of the 14′ diameter JWST-TDF, centered on 17 22 47.9, +65 49 22 (J2000). Figure 1 illustrates the SCUBA-2 area, overlaid on the field of view of other multiwavelength coverages of the JWST-TDF. While we observed simultaneously at 450 and 850 $\mu$m, the weather conditions were generally not favorable for deep 450 $\mu$m observations[12] (Figure 2). Therefore, we focus only on the deep 850 $\mu$m data in this paper.

The observation was split into 62 MSBs, each of which corresponds to a map with an on-source exposure time of $\sim$40 minutes. The total on-source exposure time is 41.3 hr (Figure 2). The sky opacity ($\tau_{225}$), representing the weather conditions of the observation, ranged from 0.02 to 0.11, with a median value of 0.09. While we were awarded Band 3 SCUBA-2 time for the 850 μm imaging, 13% and 21% of the data were obtained in better weather conditions in Band 1 ($\tau_{225} < 0.05$) and Band 2 ($0.05 < \tau_{225} < 0.08$). The observation log is summarized in Figure 2.

---

[12] The rms noise was $\sim$16 mJy beam$^{-1}$ even in the deepest region of the stacked 450 $\mu$m map using only Band 1 data. The only significant source detected is a radio-loud QSO (TDF.0005).

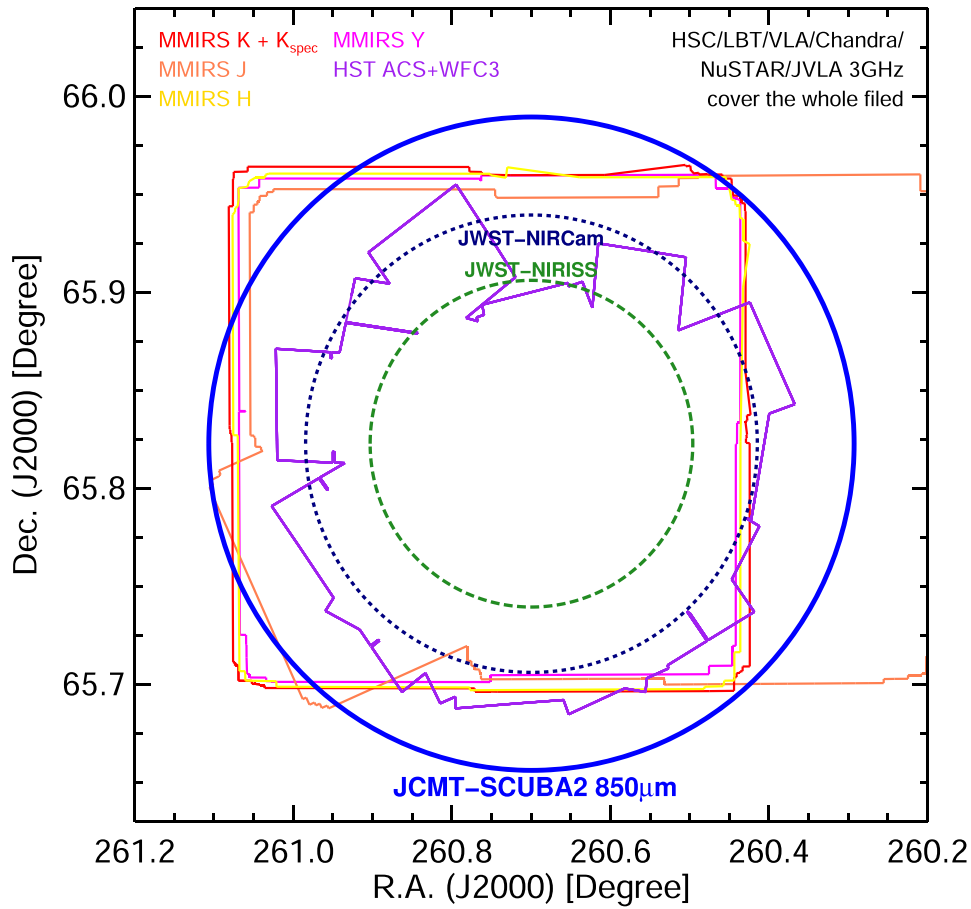

**Figure 1.** Illustration of the multiwavelength data coverage in the JWST-TDF survey region. The inner dashed green and dotted dark blue circles are the NIRISS and NIRCam survey areas, respectively (10′ and 14′ diameters, respectively). The data coverage from other facilities is shown in different colors as indicated (for more details, see Jansen & Windhorst 2018). The solid blue circle represents the JCMT SCUBA-2 850 μm survey area analyzed in this work and the VLA 3 GHz observations, with a 24′ diameter field of view, cover the full field.

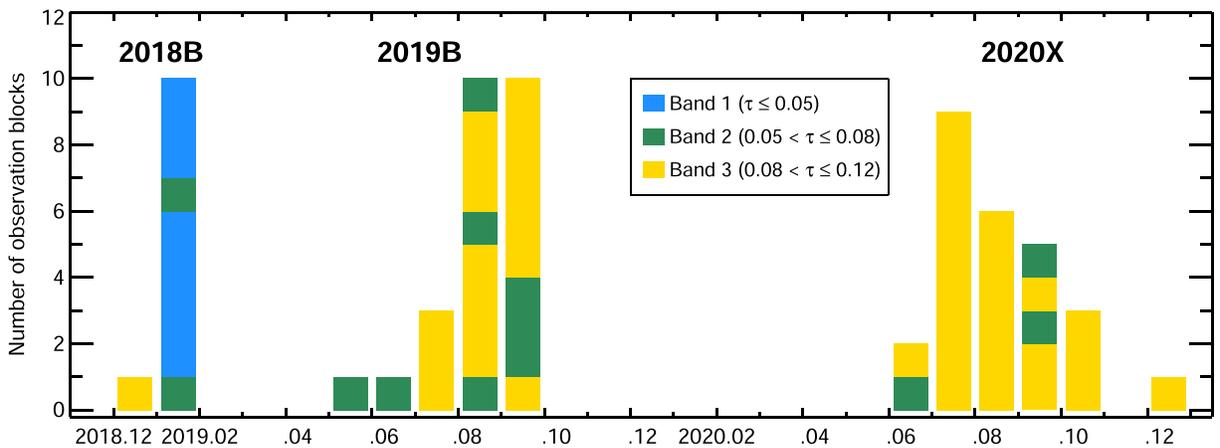

**Figure 2.** The distribution of the JCMT-TDF SCUBA-2 observations undertaken in the 2018B, 2019B, and 2020X semesters. Colors represent the weather conditions of the observations, defined in "bands" calculated from the sky opacity, $\tau_{225}$. The data for 2018B were obtained typically in better conditions than those in which the data for the 2019B and 2020X semesters were obtained. Each observation block (minimum schedulable block (MSB)) consists of an ~40 minute exposure with the PONG900 scanning mode. The total exposure time is 41.3 hr (corresponding to 62 MSBs).

### 2.2. Data Reduction

The SCUBA-2 data were reduced using the Dynamical Iterative Map Maker (DIMM) within SMURF (Sub-millimeter User Reduction Facility) with tools from the STARLINK KAPPA software package (Warren-Smith & Wallace 1993; Jenness et al. 2009) to deal with NDF format images. We utilized the "REDUCE_SCAN_FAINT_POINT_SOURCES" recipe in the ORAC-DR data reduction pipeline for a nearly automatic reduction of SCUBA-2 data using SMURF and KAPPA. For more detailed information on the full data reduction process with SMURF, see Chapin et al. (2013).

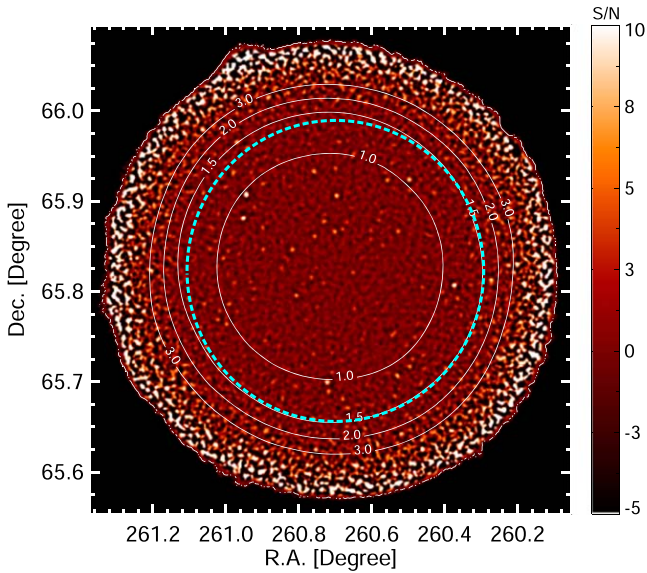

**Figure 3.** The S/N map for the 41.3 hr of SCUBA-2 observations of the JWST-TDF. The cyan dashed circle is the area ($r = 10'$) where we performed source detection ($\sigma_{inst} < 1.5$ mJy beam$^{-1}$) and the noise levels (1.0, 1.5, 2.0, and 3.0 mJy) are indicated by white contours. In the deepest region of the field, the instrumental noise reaches 0.8 mJy beam$^{-1}$, and the mean of $\sigma_{inst}$ is 1.0 mJy beam$^{-1}$.

Flux calibration was applied to convert the picowatt units in the reduced map into units of janskys. The calibrated maps were produced with the standard flux conversion factor of 537 Jy beam$^{-1}$ pW$^{-1}$ during the application of the PICARD recipes. For more information, see Dempsey et al. (2013). Finally, maps of each ∼40 minute observation were generated with 4″.0 pixel sampling.

Using all the reduced and calibrated maps, we constructed a stacked final map by applying the PICARD recipe "MOSAIC_JCM-T_IMAGES." Since our main targets are faint and compact, we used a matched filter recipe "SCUBA2_MATCHED_FILTER" to improve their detection. The matched filter process first smooths the map using a large Gaussian kernel and then subtracts the smoothed map from the original to eliminate large-scale residual noise. Then, the map was convolved with the SCUBA-2 beam.

The flux calibration of the SCUBA-2 maps was affected by the matched filtering, resulting in a ≃10% decline in flux (Simpson et al. 2019). To calibrate this factor, we inserted artificial bright sources into the map and recovered their flux after filtering. Then, we cropped the map within a radius of 600″ by using the PICARD recipe "CROP_SCUBA2_IMAGES" to isolate the region where the sensitivity is high and uniform (a mean of $\sigma_{850\mu m} = 1.0$ mJy beam$^{-1}$) to detect sources, but noting that this region entirely covers the full planned JWST-TDF footprint.

Figure 3 shows the final signal-to-noise ratio (S/N) map of the 850 μm SCUBA-2 coverage of the JWST-TDF. The noise in the deepest regions of the map reaches $\sigma_{rms} = 0.8$ mJy beam$^{-1}$, which is close to the SCUBA-2 850 μm confusion limit,[13] $\sigma_{850\mu m} \sim 0.7$ mJy beam$^{-1}$. Figure 4 presents the cumulative area of the SCUBA-2 field as a function of instrumental noise. We plot the corresponding values for the S2CLS EGS (Zavala et al. 2017), the S2CLS GOODS-N (Geach et al. 2017), the SCUBA-2 survey in the COSMOS field (Casey et al. 2013), SUPER GOODS-N (Cowie et al. 2017), SUPER GOODS-S (Cowie et al. 2018), the S2CLS NEP (Geach et al. 2017), the S2COSMOS (Simpson et al. 2019), and the NEPSC2 (Shim et al. 2020), which have a variable survey area and depth, on the figure for comparison.

---
[13] There have been several values of the confusion limit derived from various studies; however we adopted the value 0.7 mJy beam$^{-1}$, which is presented on the official JCMT website: https://www.eaobservatory.org/jcmt/instrumentation/continuum/scuba-2/time-and-sensitivity/.

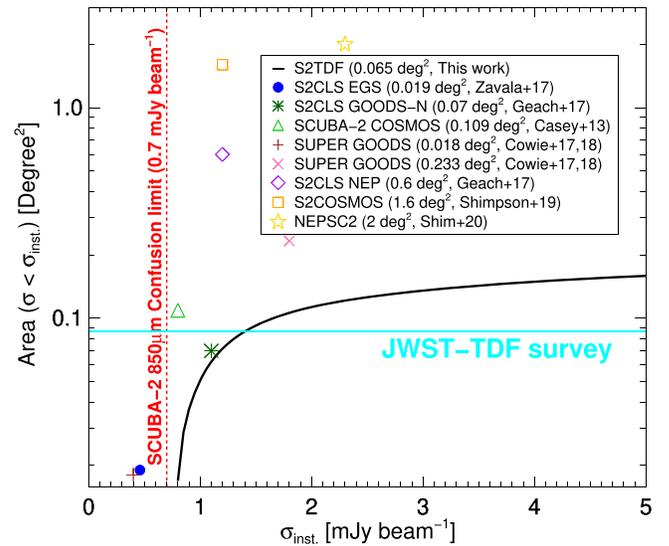

**Figure 4.** Cumulative area as a function of sensitivity of the S2TDF (black solid line). The instrumental sensitivity of our map reaches $\sigma_{inst} = 0.8$ mJy beam$^{-1}$ in the deepest region of the field covering 0.087 deg$^2$ and this is quite close to the confusion limit of SCUBA-2 850 μm, 0.7 mJy beam$^{-1}$ (red dashed line). For comparison, the S2CLS EGS (blue dot; Zavala et al. 2017) covering 0.019 deg$^2$ with a 1σ depth has $\sigma_{inst} = 0.46$ mJy beam$^{-1}$; the S2CLS GOODS-N (green asterisk; Geach et al. 2017) covering 0.07 deg$^2$ with a 1σ depth has $\sigma_{inst} = 1.1$ mJy beam$^{-1}$; the SCUBA-2 survey in the COSMOS field (light green triangle; Casey et al. 2013) covering 0.109 deg$^2$ with a 1σ depth has $\sigma_{inst} = 0.8$ mJy beam$^{-1}$; SUPER GOODS (pink cross; Cowie et al. 2017, 2018) covering 0.233 deg$^2$ with a 1σ depth has $\sigma_{inst} = 1.8$ mJy beam$^{-1}$ and reaches $\sigma_{inst} = 0.4$ mJy beam$^{-1}$ in the deepest central field covering 0.018 deg$^2$ (brown cross); the S2CLS NEP (purple diamond; Geach et al. 2017) covering 0.6 deg$^2$ with a 1σ depth has $\sigma_{inst} = 1.2$ mJy beam$^{-1}$; the S2COSMOS (orange square; Simpson et al. 2019) covering 1.6 deg$^2$ with a median noise level has $\sigma_{inst} = 1.2$ mJy beam$^{-1}$; and the NEPSC2 (yellow star; Shim et al. 2020) covering 2 deg$^2$ with a median noise level has $\sigma_{inst} = 2.3$ mJy beam$^{-1}$. The horizontal cyan solid line shows the final area ($r = 600″$) of this work where we performed source detection (see Figure 3).

## 3. Source Catalog

### 3.1. Source Detection

The final SCUBA-2 flux map is shown in Figure 5. For source detection, we adopted a simple top-down peak-finding method, which is widely used in submm studies of cosmological survey fields as this approach is effective in deblending sources with a large difference in fluxes (Geach et al. 2017; Simpson et al. 2019; Shim et al. 2020). Specifically we followed the approach used by Simpson et al. (2019). In the filtered map, optimized to detect faint point sources, we searched for sources with prominent peaks in S/N in the region with $\sigma_{inst} < 1.5$ mJy beam$^{-1}$ (Figure 3) and recorded their properties (such as their positions, flux densities, etc.) in a "first-pass" catalog. After the first detection pass, we subtracted these sources from the map by modeling their emission with an empirical point-spread function (PSF) made by using all

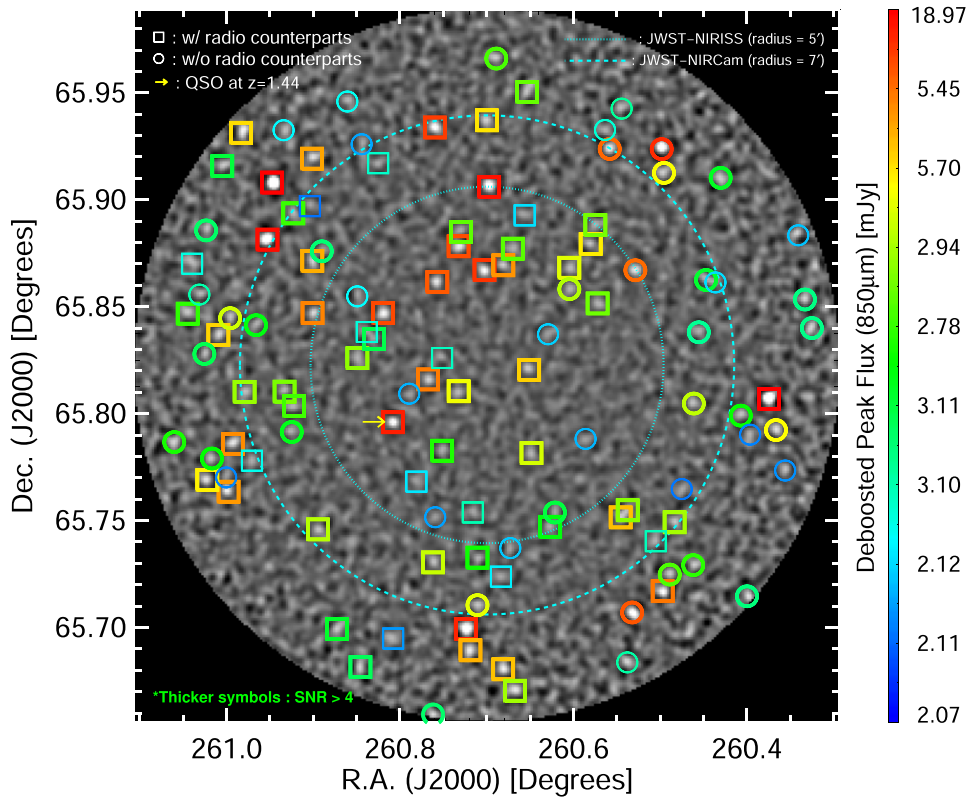

**Figure 5.** SCUBA-2 flux map of JWST-TDF showing 83 detected sources at S/N > 4 and a further 31 with S/N = 3.5–4. The color of the sources represents their deboosted flux (see color bar) and thicker symbols represent S/N > 4 sources. By matching the 850 μm sources to the VLA 3 GHz catalog using a corrected Poissonian probability estimate, we found that 66 submm sources have radio counterparts (box symbols). Circle symbols are submm sources without radio identifications. The JWST NIRISS and NIRCam survey areas are indicated by dotted and dashed circles, respectively.

>5σ sources in the map. If two sources lay within 40″ of each other, we used a double-PSF model. Using the map where the "first-pass" sources had been subtracted, we repeated the detection step for a second pass. If additional sources were detected within 7″.5 of the first-pass sources, we regarded the fluxes as the same sources as the first-pass sources and recorded the information from the prior detections only. This iteration went on until all sources above a minimum threshold of 3.5σ were found. More details are given in Simpson et al. (2019).

In this manner, we detected 83 submm sources above a 4σ significance limit and a total of 114 above a 3.5σ limit in the 850 μm map. Figure 5 shows the positions of all detected sources in the field. The thicker symbols represent the sources with S/N > 4. A histogram of the fluxes of the sources is shown in Figure 6.

### 3.2. Jackknife Simulation

Jackknife simulations are a widely used method to estimate the completeness, flux deboosting rate, false detection rate, and positional accuracy of submm surveys (e.g., Scott et al. 2008; Simpson et al. 2019; Shim et al. 2020). To construct a jackknife map, we randomly divided our SCUBA-2 MSBs into two groups, constructed a combined map from each group, and subtracted one from the other to remove any real astronomical signals. Next, the map where sources had been removed was scaled down by the square root of the total integration time to match the noise level in the actual final map.

In this source-free "noise" map, we injected artificial sources based on the expected source number counts. We assumed the number density distribution followed a Schechter function with the parameters from the 850 μm number counts in Geach et al. (2017), $N_0 = 7180$, $S_0 = 2.5$, and $\gamma = 1.5$:

$$\frac{dN}{dS} = \left(\frac{N_0}{S_0}\right)\left(\frac{S}{S_0}\right)^{-\gamma} \exp\left(-\frac{S}{S_0}\right). \quad (1)$$

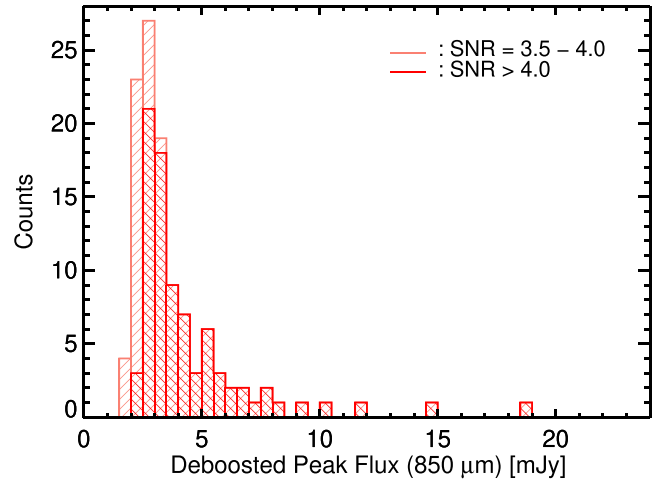

**Figure 6.** The deboosted 850 μm flux density distribution of the 114 detected SCUBA-2 sources. SNR is the S/N from the observation. Eighty-two and thirty-two sources have S/N > 4.0 (red histogram) and S/N = 3.5–4.0 (pink histogram), respectively, and every source having a deboosted flux of >2.3 mJy is in this regime.

Based on the expected number counts and our map area we inserted 500 sources into each noise map with 850 μm fluxes of 1–20 mJy, using the above number distribution. We placed the sources at random positions in the jackknife map, neglecting any clustering. We produced 1000 jackknife-simulated maps resulting in a mock catalog consisting of 500,000 sources in total. The source detection processes were the same as the process applied to the science maps described in Section 3.1. Finally, we compared the output source properties to those in the original input catalog.

### 3.3. Completeness

From the jackknife simulation, we estimated the survey completeness using the recovery success rate of injected sources. Figure 7 shows the completeness as a function of the intrinsic flux density. The 50% and 80% completeness limits are roughly 3.0 mJy and 4.0 mJy, respectively.

The completeness is not uniform across the survey area. Therefore, the completeness presented in Figure 7 is an average value over the survey area. To obtain a more accurate completeness, we derived a two-dimensional completeness map as a function of the local instrumental noise ($\sigma_{\rm inst}$) and the deboosted source flux (see Section 3.5 for the deboosting process). Here, the instrumental noise was taken as the standard deviation of the jackknife noise map at the source position. The two-dimensional completeness map is presented in Appendix B. When we measured the number counts of 850 $\mu$m sources in Section 5.1, we applied the two-dimensional completeness correction to every source.

### 3.4. False Detection Rate

At low S/N, there is a possibility that noise fluctuations will appear as spurious sources. To assess the reliability of the source catalog, we performed source detection in the same way as the real source detection on both the jackknife "noise" map and the jackknife map with artificial sources added (termed the "source" map).

By comparing the number of detections in the "noise" map and that in the "source" map, we evaluated the false detection rate as a function of S/N (Figure 7). Here, the false detection rate was defined as the ratio between the number of fake sources (false detections) and the number of detected sources. At $>3.5\sigma$, the false detection rate is $\simeq 8\%$ and it drops to $\simeq 1\%$ at $>4.0\sigma$.

A detection threshold of $>3.5\sigma$ gives a good balance between the sample size and the false detection rate for statistical studies of the SMG population (e.g., counts), while a threshold of $>4\sigma$ gives a sample with high reliability for studying their detailed properties.

### 3.5. Deboosting of Fluxes

Due to the large PSF size of the SCUBA-2 map, submm fluxes of sources, especially those near the flux limit of the survey, were often artificially boosted due to blending with positive noise peaks or faint sources. The correction of the boosted fluxes is called "deboosting."

For flux deboosting, we derived the boosting factor ($B$) from the ratio of the output (observed) flux density to the input flux density of sources in the mock catalog from the jackknife simulation. The average boosting factor at a given flux is illustrated in Figure 7, which is well fitted by a power law in S/N (Geach et al. 2017; Shim et al. 2020) as given below for our sources:

$$B = 1 + 0.3 \times \left(\frac{\rm S/N}{6.1}\right)^{-1.6}. \quad (2)$$

As noted earlier, in practice, the flux limits vary across the survey field. Therefore, we derived the boosting factor as a function of the map sensitivity as represented by the instrument noise, $\sigma_{\rm inst}$, and the flux density (before the boosting correction) as described in Appendix B. All the flux values were deboosted with the relevant boosting factors. Furthermore, we derived the additional uncertainties in the flux values that were introduced during the deboosting process (see Appendix B), and note the uncertainties as $\sigma_{\rm deb}$.

### 3.6. Positional Uncertainty of Sources

We can estimate the positional uncertainties of the detected sources from the jackknife simulation by comparing the input position with the detected (recovered) position. We derived the rms dispersion of the positional differences ($\sigma$) in the R.A. and decl. coordinates from the jackknife simulation. The $\sigma$ values were found to be inversely proportional to the S/N of the source as expected, and to have nearly identical values in both coordinates as given in the equation below:

$$\sigma = 1\rlap{.}''8 \times \left(\frac{\rm S/N}{5}\right)^{-1.0}. \quad (3)$$

Additionally, we must consider an S/N-independent intrinsic positional error ($\sigma_0$) associated with errors in the telescope pointing. This intrinsic positional error is independent of S/N and can usually be described by a Gaussian distribution with a constant rms $\sigma_0$ of bright SMGs whose $\sigma$ values are very small and negligible compared to $\sigma_0$ (Condon 1997). We estimated the intrinsic positional error by matching high-S/N SCUBA-2 sources (S/N > 10) with VLA source positions and excluding SCUBA-2 sources with multiple VLA source matches. This gave five SCUBA-2 sources, and since the number of sources is small, we used all of the R.A. and decl. offsets together to derive $\sigma_0$ assuming the intrinsic positional errors had an identical distribution. We found that the S/N-independent positional error is $\sigma_0 \sim 0\rlap{.}''9$. Considering the VLA positional errors ($0\rlap{.}''1$) are much smaller than this value, we ignored their contribution to $\sigma_0$.

In sum, the positional uncertainty in both the R.A. and decl. directions can be taken as the equation below:

$$\sigma_{\rm tot} = \sqrt{\sigma^2 + \sigma_0^2}. \quad (4)$$

In Figure 7, we show $\sigma_{\rm tot}$ in bins of S/N, with the contributions from the S/N-dependent (Equation (3)) and intrinsic ($\sigma_0$) terms indicated.

To check the consistency between the positional uncertainty based on the data and the $\sigma_{\rm tot}$ explained above, we present the standard deviation of the R.A. and decl. positional offsets between the SCUBA-2 sources and matched VLA sources and their errors estimated from bootstrapping.

For the Gaussian distributions of R.A. and decl. errors, the radial position offset ($\rho$) has a distribution that follows the Rayleigh distribution of $\rho/\sigma_{\rm tot}^2 \exp[-\rho^2/(2\sigma_{\rm tot}^2)]$. For this distribution, 68% and 95.6% of the radial offsets would lie within $1.5\sigma_{\rm tot}$ and $2.5\sigma_{\rm tot}$, respectively (Condon 1997; Ivison et al. 2007; Geach et al. 2017).

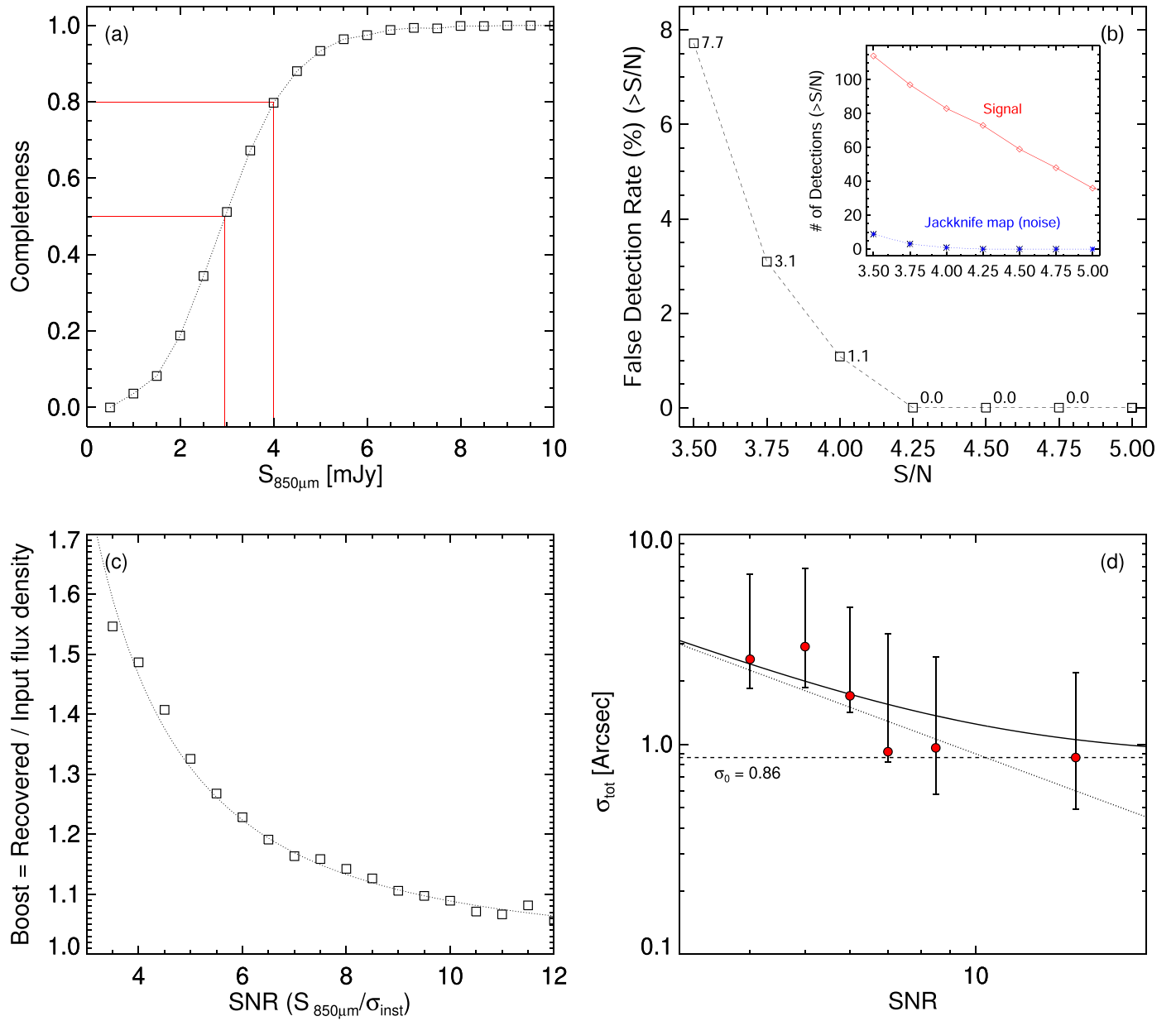

**Figure 7.** (a) Completeness of the JWST-TDF SCUBA-2 survey calculated from the ratio of the number of recovered sources with S/N > 3.5 to the number of input sources. The 50% and 80% completeness limits are 3.0 and 4.0 mJy (deboosted flux), respectively. (b) The false detection rate, the ratio between the number of sources detected in the jackknife map and that of sources detected in the flux density map (inner panel), as a function of S/N. With an S/N > 3.5 cut the probability that a detected source is spurious is ≃8%, and at S/N > 4 it is ≃1%; this drops to 0% at S/N > 4.25. (c) Average flux boosting factor as a function of S/N, showing that the results are well fitted with a power law. However, we evaluated boosting rates for each source based on the observed flux density and the local instrumental noise by constructing a two-dimensional parameter space. (d) The total positional error in both R.A. and decl. directions (solid line), which is the quadratic sum of the intrinsic positional error ($\sigma_0$, dashed line) and the positional difference with the bootstrapping error ($\sigma$, dotted line), as a function of S/N. Red filled circles are the standard deviation of the R.A. and decl. position offsets between the SCUBA-2 sources and their VLA counterparts.

### 3.7. Astrometric Calibration Using VLA

During JCMT observations, the calibration process checks and revises the telescope's pointing, and usually, the values of the shift are of the order of ~1″. However, it is important to determine precise absolute positions if we wish to identify the counterparts to the submm sources and derive their properties. Even a slight offset will make a significant difference in the calculation of the matching probability with other higher-resolution data. This astrometric calibration is typically achieved using radio maps, for example those from the VLA, with arcsecond resolution that is tied to the FK5 to a better than ~0″.01 absolute astrometric precision (e.g., Ivison et al. 1998, 2002).

We used the sources in the VLA 3 GHz catalog of this field described in Appendix A to improve the absolute astrometry of our JWST-TDF SCUBA-2 map. From the radio source catalog, detecting 756 sources at S/N > 5, we selected sources with a 3 GHz flux density of ⩾10.0 μJy, resulting in 557 VLA sources across our survey area. We used these to search for counterparts to our submm sources using a matching radius of 7″.5.[14] We identified possible radio counterparts for 68 submm sources. To determine the astrometric offset of the SCUBA-2 positions with respect to the radio source positions, we then stacked thumbnails of the SCUBA-2 sources centered on the matched radio source positions and made a $74'' \times 74''$ stacked image (Figure 8). Through this procedure, we found a small mean coordinate shift for the SCUBA-2 map of $0''\!.83 \pm 0''\!.22$ and $-0''\!.34 \pm 0''\!.03$ in R.A. and decl., respectively. We corrected our cataloged SCUBA-2 source positions for this offset. We also tested the sensitivity of the derived offset to the adopted pixel scale of the SCUBA-2 map by repeating this process with data regridded to finer pixel scales of $0''\!.5$ and $2''\!.0$ sampling, finding no significant differences.

---
[14] We set the maximum matching radius to 7″.5, which is about half of the JCMT SCUBA-2 effective beam FWHM ($\theta_{1/2} = 14″.6$; Dempsey et al. 2013) at $\lambda = 850$ μm. This has been widely used as a matching radius for radio identification (e.g., Shim et al. 2020; Zhang et al. 2022), corresponding to ≃3 times the expected total positional error (2″.4 for a 4.0σ detection and 2″.7 for a 3.5σ detection) with a 14″.6 FWHM (see Equation (4)).

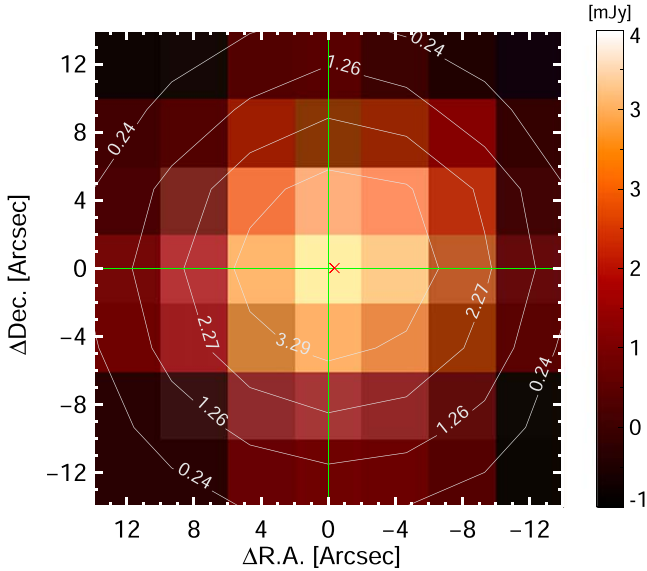

**Figure 8.** A $28'' \times 28''$ zoomed-in image of the stacked SCUBA-2 850 μm image at the positions of the radio counterparts matched to 67 SCUBA-2 sources. The red cross is the center from the two-dimensional Gaussian fitting of each stacked image. This shows a small mean coordinate shift of $0''\!.83 \pm 0''\!.22$ and $-0''\!.34 \pm 0''\!.03$ in R.A. and decl., respectively, which we corrected for in our SCUBA-2 catalog positions.

### 3.8. JWST-TDF 850 μm Source Catalog

We constructed two catalogs of the SCUBA-2 sources in the JWST-TDF (Table 4 in Appendix D) for the main (S/N > 4) and supplementary sources (S/N = 3.5–4.0), comprising eighty-three and thirty-one 850 μm sources, respectively. The coordinates of the sources include the astrometric corrections derived in Section 3.7.

There are three different flux uncertainties given in the catalog: the instrumental noise ($\sigma_{\rm inst}$), the deboosting uncertainty ($\sigma_{\rm deb}$), and the total uncertainty ($\sigma_{\rm total}$). The total flux uncertainty is the square root of the squared sum of the instrumental noise, the deboosting uncertainty, and the confusion noise ($\sigma_c$) resulting from faint sources below the detection flux limit within the JCMT beam. We calculated the confusion noise using Equation (3) of Simpson et al. (2019):

$$\sigma_c = \sqrt{\sigma_{\rm total}^2 - \sigma_{\rm inst}^2} \quad (5)$$

where $\sigma_{\rm total}$ and $\sigma_{\rm inst}$ are derived from the standard deviation of the 850 μm map where sources over the detection limit were removed, and from that of the jackknife map, respectively. The confusion noise in this study from Equation (5) is 0.14 mJy beam$^{-1}$. The S/N of the sources were not deboosted.

## 4. Band Merging and SED Analysis

### 4.1. Multiwavelength Matching

We matched the detected 850 μm sources in the JWST-TDF to radio counterparts from the 3 GHz catalog described in Appendix A. These radio counterparts are essential to providing the precise positions necessary to determine the multi-wavelength properties of the likely SMGs from the optical to radio. For SMGs with radio counterparts and optical/NIR photometric detections, we derived their redshifts and other properties through SED fitting.

#### 4.1.1. Radio Counterparts

To reliably identify the multiwavelength counterparts to the submm sources, we exploited the correlation between the radio and FIR luminosities of normal or star-forming galaxies (e.g., Rickard & Harvey 1984; Helou et al. 1985; Condon 1992; Ivison et al. 2004, 2005, 2008), which has been widely used in previous submm studies (e.g., Barger et al. 2000; Smail et al. 2000; Ivison et al. 2007; An et al. 2019). We first crossmatched the SCUBA-2 sources with our VLA 3 GHz radio catalog (see Appendix A). For matching, we used the 557 radio sources with 3 GHz flux densities above the uniform limit of $\geqslant 10$ μJy (as used in Section 3.7).

The corrected Poissonian probability, $p$, used in the matching analysis was defined as follows:

$$p = 1 - \exp(-E) \quad (6)$$

$$E = \begin{cases} P_c & P^* \geqslant P_c \\ P^* \{1 + \ln(P_c/P^*)\} & P^* \leqslant P_c \end{cases} \quad (7)$$

$$P_c = \pi r_s^2 N_T \quad (8)$$

$$P^* = \pi r_i^2 N_i. \quad (9)$$

In line with previous work we adopted a maximum search radius (see Section 3.7), $r_s$, of $7''\!.5$. The variable $r_i$ is the angular distance between the submm centroid and the matching radio source (with 3 GHz flux $S_i$). $N_T$ and $N_i$ are the radio source number densities, with $S_{\rm 3GHz} > 10$ μJy and $S_{\rm 3GHz} > S_i$, respectively. For each SCUBA-2 source, we calculated the probability, $p$, for any radio sources within $r_s$, and retained only sources having radio counterparts with $p \leqslant 0.065$ based on the balance of recovery and false-positive rates in the analysis of an Atacama Large Millimeter/submillimeter Array (ALMA)-identified sample of SMGs by An et al. (2018).

Figure 9 summarizes the results of the radio identifications. The SCUBA-2 sources with radio identification are listed in Table 4 along with the corresponding probabilities. Note that the number after the decimal point in the source ID indicates the radio counterpart identifications, ranked on the probability of the match; it can be larger than 0 if there are multiple matches. Using a maximum search radius of $7''\!.5$, there are 89 radio identifications. Four radio sources have likelihoods of being false matches of $p > 0.065$, and removing these reduces the total number of robust matches to 85 radio sources. There are 52 submm sources with single radio counterparts, nine with two radio counterparts, and five with three counterparts. The SCUBA-2 sources with multiple radio counterparts have an 850 μm flux density distribution that is indistinguishable from that of the parent population of all SCUBA-2 sources with radio counterparts ($S_{850\mu m} = 3.3 \pm 0.3$ mJy versus $S_{850\mu m} = 3.6 \pm 0.7$ mJy), while the fluxes of the SCUBA-2 sources lacking radio counterparts are somewhat fainter on average: $S_{850\mu m} = 2.8 \pm 0.2$ mJy. For the main sample, 54 out of 83 SCUBA-2 sources matched with 70 radio sources including five/six triple/double-counterpart cases. For the supplementary sample, 12 out of 31 SCUBA-2 sources matched with 15 radio sources, among which there are three double-counterpart cases and no triple-counterpart cases.

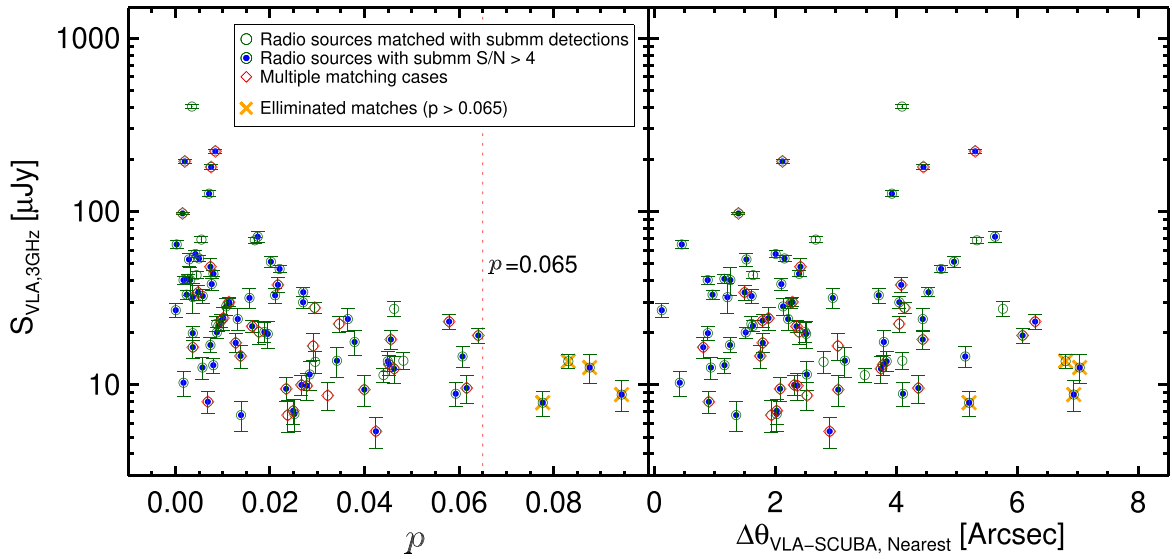

**Figure 9.** VLA 3 GHz flux densities of 89 radio sources matched to the 68 SCUBA-2 source positions, as a function of corrected Poissonian probability, $p$ (left), and as a function of the angular separation between the radio sources and the SCUBA-2 sources (right). SCUBA-2 sources with S/N > 4 are marked with blue filled circles. There are 37 multiple radio counterpart matches to 16 SCUBA-2 sources, and these are indicated with red diamond marks. After the probability cut $p = 0.065$ is applied, 33 multiple matches to 14 SCUBA-2 sources remain. The red vertical line is the probability cut adopted from the analysis of ALMA identifications of SMGs by An et al. (2019). With that criterion, four radio sources are removed from the radio-identified sample. Hence, in total 85 radio sources are matched to 66 SCUBA-2 sources.

To summarize, 85 VLA radio sources are matched to 66 of the 114 SCUBA-2 sources, and these matches are typically sources that are at the brighter end of the 850 $\mu$m flux distribution. The overall matched fraction corresponds to an identification rate of 58% (65% and 39% for the main and the supplementary sample), which is broadly consistent with that seen in other radio-identified SMG samples at these depths (60%–80%, e.g., Ivison et al. 2007; Casey et al. 2013; Miettinen et al. 2015).

#### 4.1.2. Optical/NIR Counterparts

For those SCUBA-2 sources with radio counterparts, we used the precise radio positions and our multiwavelength coverage to characterize the properties of these galaxies. We used the radio positions and matched these to the ground-based band-merged optical and NIR catalog of the JWST-TDF (C. N. A. Willmer et al. 2022, in preparation) with a search radius of $1\rlap.{''}0$. This radius is small enough that we expected no false matches.

The band-merged catalog comprises optical photometry in the $g$, $i$, and $z$ bands based on source detections in the combined $i+z$ image (appropriate for detecting red galaxies such as SMGs). These data were obtained with Hyper Suprime-Cam (HSC; Bosch et al. 2018; Miyazaki et al. 2018), on the Subaru telescope, and have $3\sigma$ aperture magnitude limits of $g = 26.0$, $i = 25.2$, and $z = 25.1$ in $3\rlap.{''}0$ diameter apertures. The $3\rlap.{''}0$ aperture magnitude is roughly identical to total magnitudes for compact/faint galaxies.

This optical catalog has been merged with NIR catalogs in the $Y$, $J$, $H$, and $K$ bands from the Magellan Infrared Spectrograph (MMIRS; McLeod et al. 2012) on the Multi-Mirror Telescope (MMT). The $3\sigma$ aperture magnitude limits of the MMIRS images are $Y = 24.7$, $J = 23.7$, $H = 23.0$, and $K = 22.5$, again in $3\rlap.{''}0$ diameter apertures, to provide total magnitudes. This NIR coverage has been supplemented by the WISE 3.6–22 $\mu$m counterparts, matched to the radio sources using a $1\rlap.{''}0$ search radius.

There are 61 radio-identified SMGs matched to the merged optical/NIR catalog. Of these 61, 54 are detected above $3\sigma$ in the NIR (16 are detected with WISE), and the remaining seven have optical HSC detections and upper limits in the other bands.

#### 4.2. SED Fitting Analysis

We performed an SED analysis of the 61 radio-identified, optical/NIR-matched SMGs to derive both photometric redshifts and other physical properties (we similarly modeled a further 24 sources that had radio and submm constraints but no optical or NIR detections). We employed the MAGPHYS galaxy SED-fitting code (da Cunha et al. 2008; Cunha et al. 2015; Battisti et al. 2019). MAGPHYS uses an energy balance technique to combine observations from the optical/NIR, MIR/FIR, and submm wave bands to constrain the physical properties of sources, including the stellar population and dust content. The "high-redshift" version of the code, MAGPHYS+PHOTOZ, also allows the redshift of the source to be estimated (Battisti et al. 2019). The ability of the code to include information from the dust-reprocessed emission seen in the MIR/FIR and submm is important when attempting to model SMGs, which are frequently faint or undetected in the optical and NIR wave bands used by other photometric modeling codes. MAGPHYS+PHOTOZ has been used to model samples of SMGs, for example by Dudzevičiūtė et al. (2020), who provided an extensive discussion of the calibration and testing of the code on dust-obscured galaxies, including its application to modeling SEDs derived from radiative transfer models of simulated galaxies from the EAGLE simulation (see also McAlpine et al. 2019).

Our SED modeling used the optical/NIR/MIR photometry for the matched counterparts, and for the SMGs, also included our deboosted 850 μm observations and 3 GHz VLA data. In the analysis, we followed a similar approach to that used in Dudzevičiūtė et al. (2020), first applying MAGPHYS+PHOTOZ to the collated photometry of a sample of 264 sources, including four SMGs and 66 VLA sources, with robust spectroscopic redshifts (spanning $z_{spec} = 0$–1.6) in our field (C. N. A. Willmer, private communication; see Appendix C), to check for systematic errors in either the total magnitudes or the absolute calibration of the photometry. For each filter, we determined the distribution of differences between the observed and model photometry for the best-fit MAGPHYS model SEDs of the sources at their known spectroscopic redshifts. This analysis confirmed that there were no significant, >0.10 mag, systematic offsets in any of the filters, and so we proceeded to model the galaxies allowing the code to fit for the redshift. This returned a median difference between the predicted photometric and spectroscopic redshifts of $\Delta z/(1 + z_{spec}) = -0.10 \pm 0.02$ for the 264 galaxies, indicating that the photometric redshifts are unbiased, with a scaled median absolute deviation of $\sigma_z/(1 + z_{spec}) = 0.19$.

We next applied MAGPHYS+PHOTOZ to the 61 SMGs in our field that had 3 GHz radio detections and suitable optical and/or NIR photometry (and also the 25 with just radio and submm detections). The best-fit model SEDs for the sources are shown in Figure 17 in Appendix D.

MAGPHYS+PHOTOZ returns estimates of various physical quantities of the modeled galaxies, not all of which are well constrained by the observable properties (see the discussion in Dudzevičiūtė et al. 2020). Given the limited photometric coverage used in our analysis, we consider the redshifts and dust masses as the most robust output parameters, although we also discuss the estimates of the SFR and stellar mass in Sections 5.2 and 5.3.

The estimated 16th–84th percentile errors on the various physical parameters from MAGPHYS+PHOTOZ are as follows: photometric redshift, $\delta z_{phot}/(1 + z_{phot}) = 0.21$; stellar mass, $\delta M_* = 0.37$ dex; SFR, $\delta SFR = 0.35$ dex; dust mass, $\delta M_d = 0.24$ dex; and dust-to-stellar mass ratio (DSR), $\delta M_d/M_* = 0.43$ dex. In the case of the main sample (S/N > 4), the errors are slightly smaller: photometric redshift, $\delta z_{phot}/(1 + z_{phot}) = 0.22$; stellar mass, $\delta M_* = 0.36$ dex; SFR, $\delta SFR = 0.34$ dex; dust mass, $\delta M_d = 0.23$ dex; and DSR, $\delta M_d/M_* = 0.43$ dex. These uncertainties are typically $1.4 \pm 0.2$ times larger than the equivalent values derived for the 22-band coverage in AS2UDS by Dudzevičiūtė et al. (2020) reflecting the limitation of the current photometric coverage of the TDF region.

## 5. Results and Discussion

### 5.1. 850 μm Number Count

The number density of sources as a function of flux density is one of the most basic observables available for a population and as such provides a fundamental test for galaxy formation models (e.g., Baugh et al. 2005). First, we present the number counts of the submm sources detected in the JWST-TDF and compare these with those of other surveys.

**Table 1**
The Number Counts of JWST-TDF SCUBA-2

| $S_{850\mu m}$ (mJy) | $N(>S_{850\mu m})$ (deg$^{-2}$) | $S_{850\mu m}$ (mJy) | $dN/dS_{850\mu m}$ (deg$^{-2}$ mJy$^{-1}$) |
|---|---|---|---|
| 4.0  | $462.9^{+90.5}_{-77.4}$ | 4.6  | $181.6^{+65.9}_{-49.7}$ |
| 5.1  | $252.7^{+71.2}_{-54.7}$ | 5.9  | $78.6^{+40.0}_{-27.9}$ |
| 6.6  | $137.9^{+54.1}_{-42.5}$ | 7.5  | $37.0^{+24.2}_{-18.1}$ |
| 8.5  | $68.8^{+40.3}_{-27.2}$  | 9.7  | $14.3^{+13.4}_{-11.3}$ |
| 10.9 | $34.4^{+35.3}_{-21.9}$  | 12.4 | $3.7^{+9.3}_{-3.1}$ |
| 14.0 | $23.0^{+30.1}_{-18.7}$  | 15.9 | $2.9^{+7.2}_{-3.8}$ |
| 17.9 | $11.5^{+26.2}_{-14.9}$  | 20.5 | $2.3^{+5.2}_{-2.9}$ |

**Note.** Uncertainties were derived from the standard deviation in simulations for each bin using the source deboosting and completeness corrections.

Table 1 provides a summary of the number counts and Figure 10 shows the cumulative and differential counts of 850 μm sources in the ~0.087 deg$^2$ area of the JWST-TDF SCUBA-2 survey. The counts use deboosted fluxes and are corrected using the completeness curve constructed in Section 3.3. The errors reflect the uncertainties from the deboosting process, which were derived from constructing 1000 mock catalogs of 850 μm sources by sampling the deboosted flux distributions for the sources and adopting the 16th and 84th percentiles of the resulting distribution as the uncertainties. The final number counts were derived from the mean values of these mock catalogs (Table 1). We also plot the best Schechter function fits to the number counts in the AS2COSMOS (Simpson et al. 2020), S2COSMOS (Simpson et al. 2019), and S2CLS (Geach et al. 2017) surveys for comparison.

The JWST-TDF number counts in both panels trace the Schechter function–like trend well, and agree with the results from other surveys (e.g., Chen et al. 2013; Karim et al. 2013; Hsu et al. 2016; Geach et al. 2017; Stach et al. 2018; Simpson et al. 2019, 2020; Shim et al. 2020), especially at fainter flux densities ($S_{850\mu m} < 10$ mJy). There is a slight excess at $S_{850\mu m} > 10$ mJy. We believe that this is likely caused by cosmic variance and reflects four bright sources with fluxes exceeding $S_{850\mu m} = 10$ mJy in our relatively small survey area (~0.087 deg$^2$). To quantify this, we randomly placed TDF-sized regions in the much larger S2COSMOS survey field (Simpson et al. 2019) and found that regions containing ⩾5 SCUBA-2 sources brighter than $S_{850\mu m} = 10$ mJy were detected ~3% of the time—suggesting that the excess can be explained by cosmic variance.

### 5.2. Redshifts

We found that the best-fit photometric redshifts for the radio-identified and optical/NIR-detected SMGs with S/N > 4.0 (>3.5) in our sample give a median redshift of $z = 2.22 \pm 0.12$ ($z = 1.96 \pm 0.08$) and a 16th–84th percentile range of $z = 1.50$–3.10 ($z = 1.44$–2.70). We note that there are high-quality spectroscopic redshifts for only four of the dust-selected SMGs (the low rate likely reflecting the relative faintness of the galaxies in the optical/NIR). Nevertheless, in three of these four cases, the spectroscopic redshifts and photometric redshift estimates appear to agree with each other (the exception is source 0056.0, with $z_{spec} = 0.03$, a spectroscopic redshift that likely refers to a nearby foreground galaxy): 0023.0, $z_{spec} = 1.51$, $z_{phot} = 1.55^{+0.25}_{-0.27}$; 0031.0, $z_{spec} = 1.36$, $z_{phot} = 1.44^{+0.05}_{-0.05}$; and 0051.0, $z_{spec} = 1.51$, $z_{phot} = 1.48^{+0.08}_{-0.11}$.

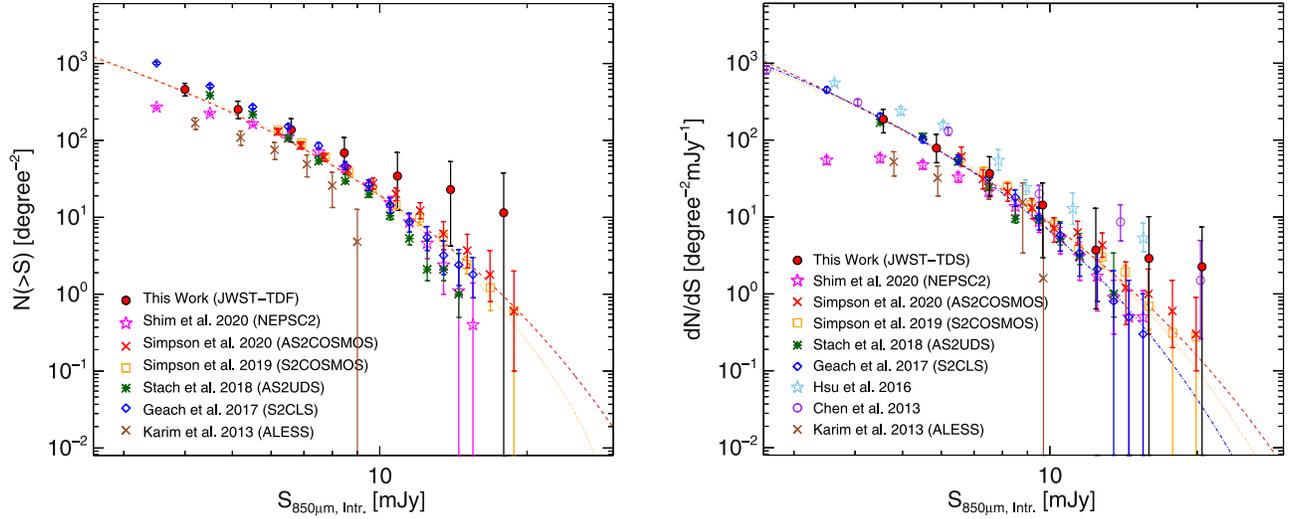

**Figure 10.** Cumulative and differential number counts of the 850 μm 114 sources ($S_{850\mu m}$ = 2–20 mJy, S/N > 3.5) in the JWST-TDF area. For comparison, we show results from previous submm surveys (Chen et al. 2013; Karim et al. 2013; Hsu et al. 2016; Geach et al. 2017; Stach et al. 2018; Simpson et al. 2019, 2020). The best-fitting results from Simpson et al. (2019, 2020), Shim et al. (2020), and Geach et al. (2017) are also presented. In general, our number counts are in reasonable agreement with those from the other surveys, although there is a slight excess at the bright end, $S_{850\mu m}$ > 10 mJy, arising from a small number of bright sources in the field.

The redshift distribution of our radio-identified 850 μm sources is presented in Figure 11. For comparison, we show the photometric redshift distribution for the purely submm-selected AS2UDS survey from Dudzevičiūtė et al. (2020), and also the equivalent distribution after applying an additional radio detection and $K$ < 22.5 magnitude limit (to mimic the selection of our radio-identified, mostly NIR-detected sample). We also show the spectroscopic redshift distribution for the similarly radio-identified SMG sample from Chapman et al. (2005). In contrast to the flux-limited AS2UDS sample, which shows a tail of sources out to $z \sim 6$, both of the radio/NIR-limited samples exhibit a sharp cutoff at $z \sim 4$ and median redshifts at $z \sim 2$, in reasonable agreement with those in the JWST-TDF.

Figure 12 shows the $S_{850\mu m}$ and $S_{850\mu m}/S_{3GHz}$ versus the photometric redshifts of the 51 (61) submm sources at S/N > 4.0 (S/N > 3.5). Also plotted are the $K$-band and radio limited sample of AS2UDS, mimicking the selection of the S2TDF sources, and the median redshifts of the submm flux-limited sample of AS2UDS. We conclude that the S2TDF sources follow the distribution of those in AS2UDS when a similar radio/NIR selection is applied to them.

### 5.3. SFRs and Stellar Masses

The distribution of SFRs with stellar mass for those S2TDF radio-identified SMGs with optical/NIR counterparts is plotted in Figure 13. The S2TDF sample shows a median SFR of $300 \pm 40 \, M_\odot \, \mathrm{yr}^{-1}$ and $240 \pm 40 \, M_\odot \, \mathrm{yr}^{-1}$ for SCUBA-2 sources with S/N > 4.0 and S/N > 3.5, respectively. The maximum SFR is $\sim 3000 \, M_\odot \, \mathrm{yr}^{-1}$ for both cases.

These galaxies have stellar masses with a 16th–84th percentile range of $0.8$–$4 \times 10^{11} M_\odot$, and a median of $1.7 \pm 0.3 \times 10^{11} M_\odot$, showing that most of them are already very massive. Adding the supplementary subset associated with SCUBA-2 sources with S/N = 3.5–4.0, we found their stellar masses are very similar with a median $M_* = 1.6 \pm 0.3 \times 10^{11} M_\odot$ and a 16th–84th percentile range of $0.6$–$5 \times 10^{11} M_\odot$. With their high SFRs, they are likely to become massive early-type galaxies today, as suggested by other SMG studies (Lilly et al. 1999;

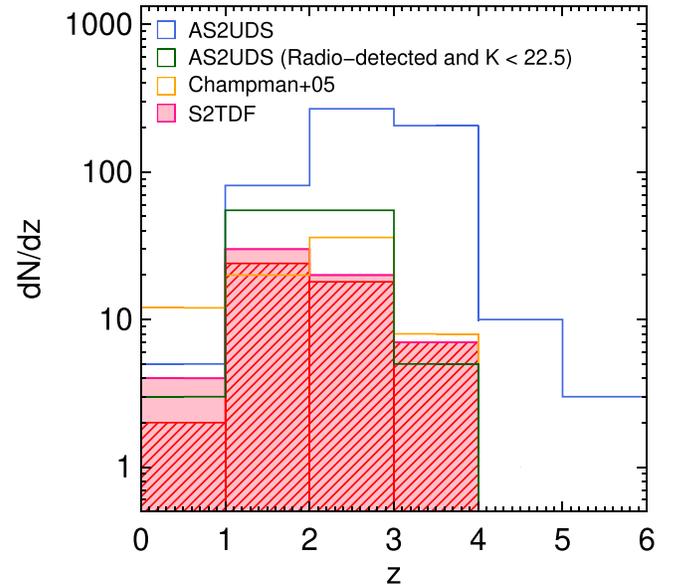

**Figure 11.** The photometric redshift distribution of radio-identified SMGs in the JWST-TDF. The red hatched histogram shows the distribution of samples at S/N > 4.0, yielding a median redshift of $z = 2.22 \pm 0.12$ and a 16th–84th percentile range of $z = 1.50$–$3.11$. If we include the sample with 850 μm S/N = 3.5–4 (pink shaded histogram), then we derive a median redshift of $z = 1.96 \pm 0.08$ and a 16th–84th percentile range of $z = 1.44$–$2.70$. The redshift distributions from the submm flux-limited survey (AS2UDS, blue histogram), the AS2UDS sample with radio detection and $K$ < 22.5 mag limits applied to mimic the selection of our sample (green histogram), and finally the radio-identified SMG sample (yellow histogram) from Chapman et al. (2005) are also shown for comparison.

Birkin et al. 2021). Figure 13 also compares the SFR–$M_*$ distribution of the S2TDF SMGs to that of the $K$-band and radio limited subsample of AS2UDS, which shows a similar distribution to the S2TDF SMGs. To compare the distribution to the trend in more typical "main-sequence" star-forming galaxies, we also show the SFR–$M_*$ relation for these galaxies in Figure 13. This comparison sample was constructed by matching the redshifts and stellar masses of the S2TDF SMGs to those of their nearest neighbors in $z$–$M_*$ space from the large sample of $K$-detected field galaxies in the UKIRT Infrared Deep Sky Survey Ultra Deep Survey (UDS) from Dudzevičiūtė et al. (2020). This matching perturbed each SMG 10 times by its uncertainties in redshift and stellar mass and selected the closest match from the field sample within a redshift window of $\delta z = 0.1$. Then, these UDS galaxies were plotted on the SFR–$M_*$ plane, and the best-fit linear correlation was derived, which is given in Equation (10):

$$\log(\mathrm{SFR}) = 0.57 \log(M_*/M_\odot) - 4.55. \qquad (10)$$

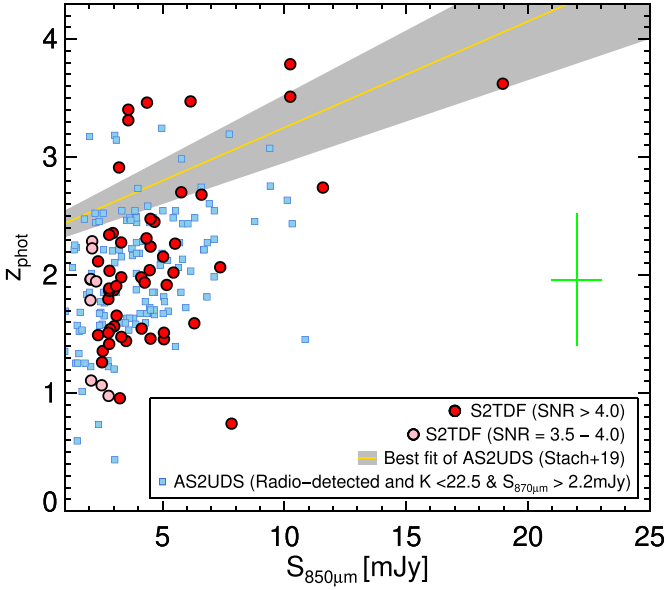
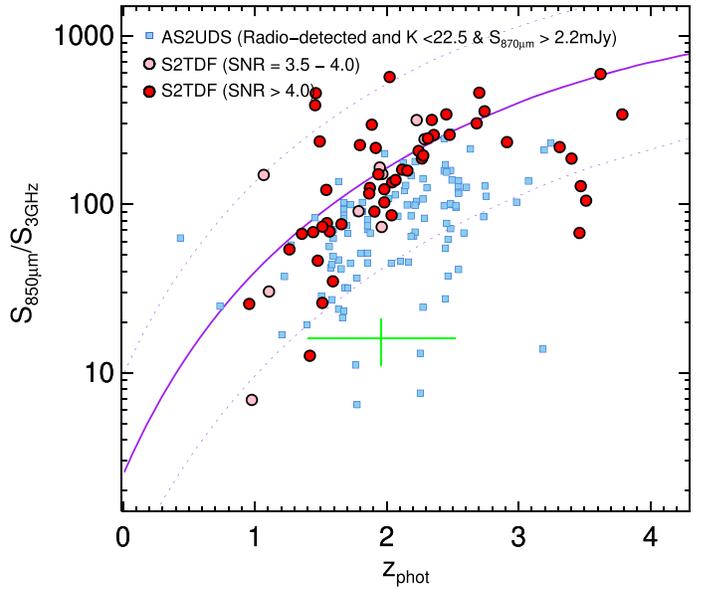

**Figure 12.** Photometric redshift vs. flux density and flux ratio for SCUBA-2 SMGs. S2TDF sources at S/N > 4.0 and at S/N = 3.5–4.0 are presented with filled red circles and with filled pink circles, respectively. The median error of each parameter is drawn in each panel (green). (Left) 850 μm flux density vs. photometric redshift for our radio-identified SMGs with optical/NIR counterparts. For comparison we show a radio and $K$-band limited subsample of SMGs from AS2UDS (Dudzevičiūtė et al. 2020), and the fitted trend for the full AS2UDS survey from Stach et al. (2019). There is broad agreement between the redshift–flux distributions in the JWST-TDF and these previous works. (Right) The variation in submm-to-radio flux ratio and photometric redshift for our sample, again compared to a radio/$K$-band-limited subset of the SMGs in AS2UDS (converted from 1.4 to 3 GHz assuming a radio spectral index of −0.7). We also show the expected trend for the composite SMG SED from Dudzevičiūtė et al. (2020), which broadly reproduces the behavior we see.

This best-fit trend for the field population is shown in Figure 13 and we compared it to the SFR–$M_*$ relations of "main-sequence" star-forming galaxies at $z = 1$–5 from Pearson et al. (2018), finding that it corresponds well to the track for $z \sim 2$. This shows that our SMG sample have higher SFRs than "typical" galaxies of their mass at their respective redshifts, by a factor of ∼6, indicating that our SMGs are examples of the most massive, highest-SFR end of the galaxy populations at $z \sim 1$–3.

### 5.4. Dust Mass Estimates

Finally, we take advantage of the fact that the 850 μm continuum brightness is tightly correlated to the cold dust mass of galaxies across a wide range of redshifts. In turn, the cold dust mass of galaxies is believed to be a good indicator of the mass of their cool gas reservoirs (e.g., Scoville et al. 2016). We derived estimates of the cold dust masses of our SMGs from the MAGPHYS fits and compared these to those for the similar radio-detected and $K < 22.5$ sample of SMGs from AS2UDS (Dudzevičiūtė et al. 2020) in Figure 14. Both samples show similar ranges in dust mass, $M_d \sim 2$–$20 \times 10^8 M_\odot$, and the S2TDF sample shows a median dust mass of $5.9 \pm 0.7 \times 10^8 M_\odot$ ($M_d = 4.4 \pm 0.8 \times 10^8 M_\odot$ when including the supplementary S/N = 3.5–4.0 subset) suggesting typical gas masses of $M_g \sim 2$–$3 \times 10^{10} M_\odot$, for gas-to-dust mass ratios of ∼60 (Birkin et al. 2021).

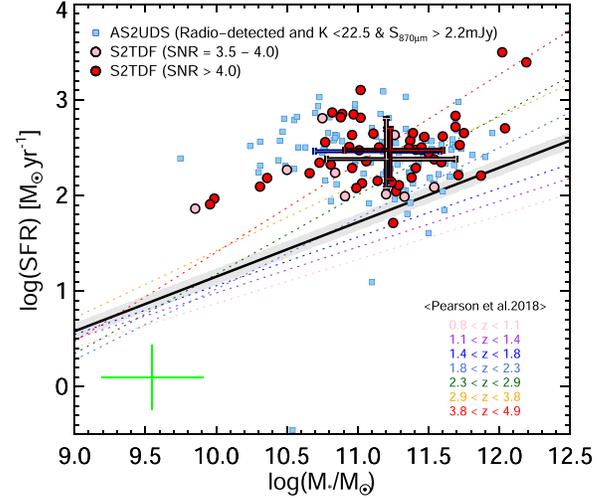

**Figure 13.** SFR–$M_*$ distribution for the S2TDF, compared to a $K < 22.5$ limited, radio-selected subsample of SMGs from AS2UDS (Dudzevičiūtė et al. 2020). We show the median error at the bottom left corner in green. The solid line is the linear trend derived from a sample of $K$-band-detected field galaxies that has been matched to the S2TDF SMGs in redshift and stellar mass (see the main text), and the shaded area indicates the 20% uncertainty in the normalization. We also show the SFR–stellar mass relation for the "main sequence" of star-forming galaxies at various redshifts derived in Pearson et al. (2018). Both of these demonstrate that the S2TDF SMGs have SFRs above those typical for similar-mass field galaxies at their redshifts.

In Figure 14, we also compare the cold dust masses for our SMGs with those inferred from MAGPHYS SED fitting to more typical "main-sequence" galaxies, using the $K$-band "field" selected sample (with a $K < 22.5$ cut) from Dudzevičiūtė et al. (2020). The median dust mass for these galaxies is $M_d \lesssim 10^8 M_\odot$, considerably lower than that of the SMGs, suggesting a similar difference in their cool gas masses. Such a difference would then provide a natural explanation for the much higher levels of star formation activity seen in the SMGs, as a consequence of their much more massive reservoirs of cool gas.

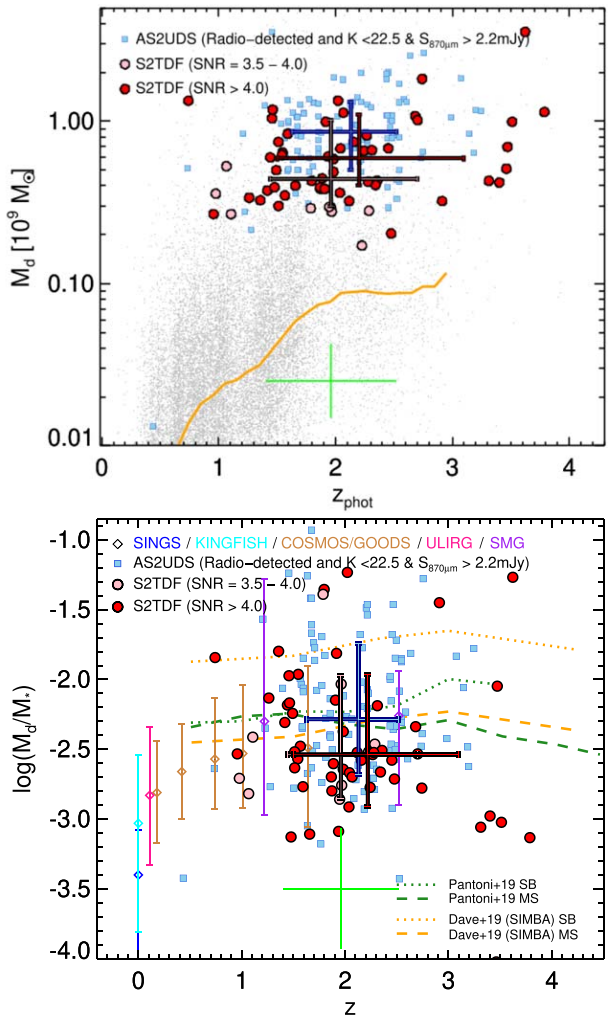

**Figure 14.** In both panels, red and pink filled circles show the S2TDF sources at S/N > 4.0 and at S/N = 3.5–4.0, respectively. The median errors are also presented (green). (Top) The variation in cold dust mass from MAGPHYS for our SMGs with redshift. We compare this to that of a similarly radio-detected and $K < 22.5$ sample of SMGs from AS2UDS (Dudzevičiūtė et al. 2020). The details of the symbols are the same as those in Figure 12. In addition, we show a $K$-band-selected field galaxy sample with $K < 22.5$ (gray points) and the trend (yellow solid line) of estimated dust masses from Dudzevičiūtė et al. (2020). This illustrates significantly higher cold dust masses, and by implication higher cold gas masses, for the SMGs, compared to typical star-forming galaxies at similar redshifts. (Bottom) DSR as a function of redshift for radio-identified SMGs in this work and in AS2UDS (Dudzevičiūtė et al. 2020). For comparison we present the median DSR of various samples collated by Calura et al. (2017) with open diamond symbols: blue, cyan, deep pink, brown, and purple colors represent spiral galaxies from SINGS (Kennicutt et al. 2003), spiral galaxies from KINGFISH (Kennicutt et al. 2011), a local ULIRG sample (Clements et al. 2010), FIR-selected galaxies in the COSMOS/GOODS surveys (Calura et al. 2017; Lutz et al. 2011), and higher-redshift SMGs (Greve et al. 2005; Santini et al. 2010), respectively. We also show the model predictions of "main-sequence" and starburst dusty star-forming galaxies from Davé et al. (2019) and Pantoni et al. (2019), respectively.

The ratio of dust mass to stellar mass in galaxies is a potentially useful parameter that links the growth of stellar populations in galaxies with the associated formation of metals in stars and dust in their atmospheres and the broader interstellar medium (ISM). Critically this ratio may provide constraints on the processes that destroy dust grains within the ISM (e.g., Santini et al. 2010; Dunne et al. 2011; Calura et al. 2017; Donevski et al. 2020; Dudzevičiūtė et al. 2021).

We show in Figure 14 the variation in the DSR for our SMGs and the radio/$K$-band-matched AS2UDS sample from Dudzevičiūtė et al. (2020). We also show the median DSRs for local and high-redshift galaxies. These comprise low-redshift spirals and ULIRGs, higher-redshift FIR-selected galaxies, SMGs, and a small number of very-high-redshift QSOs (Kennicutt et al. 2003, 2011; Greve et al. 2005; Clements et al. 2010; Santini et al. 2010; Lutz et al. 2011; Calura et al. 2014, 2017).

For the S2TDF main sample, the SMG DSR is $\log_{10}(M_d/M_*) = -2.5 \pm 0.1$ ($-2.5 \pm 0.2$, S/N > 3.5), compared to $-2.3 \pm 0.1$ for the AS2UDS matched sample, with a Kolmogorov–Smirnov two-sample test showing that the two populations are indistinguishable, $P_{KS} = 0.6$. Our median estimates agree well with earlier crude estimates for SMGs (Greve et al. 2005; Santini et al. 2010) and appear to reflect a gradual increase in DSR from $z = 0$ to $z \sim 2$ (which may in turn reflect the rising gas fraction in galaxies over this period, e.g., Geach et al. 2005; Tacconi et al. 2018; Dudzevičiūtė et al. 2020, 2021; Birkin et al. 2021).

We also plot two sets of theoretical models on Figure 14 for the "main-sequence" and "starburst" models from Davé et al. (2019) and Pantoni et al. (2019). The former cosmological models by Davé et al. (2019) suggest little variation in the DSR as a function of SFR/$M_*$ (which distinguishes the two model families), but these fail to reproduce the range in DSR: our SMGs span a 16th–84th percentile range of DSRs of $\log_{10}(M_d/M_*) \sim -2.0$ to $-2.9$, which is approximately twice the estimated uncertainty, indicating a possible intrinsic dispersion in the DSR of the population. In contrast, the simple analytic models from Pantoni et al. (2019) do broadly span the range in DSR in the SMGs. This likely reflects the relative paucity of large numbers of rapidly evolving massive, gas- and dust-rich galaxies at high redshifts in typical cosmological simulations.

## 6. Summary and Conclusion

We have conducted a SCUBA-2 850 $\mu$m survey of the JWST-TDF, which will be a prime deep extragalactic field benefiting from deep multi-epoch observations with the James Webb Space Telescope. The main results of our study are as follows.

Our JCMT SCUBA-2 850 $\mu$m survey of the JWST-TDF identified 83 and 31 submm sources, at S/N > 4 and S/N = 3.5–4, respectively, in a survey area of $\sim 0.087$ deg$^2$. The mean 1$\sigma$ sensitivity across the survey area is 1.0 mJy beam$^{-1}$ from the 41.3 hr of observation, with the deepest region achieving 0.8 mJy beam$^{-1}$. The latter is comparable to the 850 $\mu$m confusion limit of the JCMT ($\sigma_c = 0.7$ mJy beam$^{-1}$).

To derive the survey completeness and the boosting factor, we performed jackknife simulations using number counts reflecting the expected source density in the survey. The 50% and 80% completeness limits are 3.0 mJy and 4.0 mJy, respectively. The false detection rates derived from the simulations are 1% and 8% at S/N > 4.0 and S/N > 3.5. The purity of the sample approaches 100% at S/N > 4.25.

We also analyzed sensitive new VLA 3 GHz observations of the JWST-TDF. These observations have a synthesized beam of a 0″.7 FWHM and cover a 24′ field of view, reaching a 1$\sigma$ sensitivity of 1 $\mu$Jy at the field center. We have presented a catalog of 756 sources detected above a flux density limit of S/N = 5 in this field.

The SCUBA-2 sources were crossmatched with the radio sources from our deep VLA 3 GHz map. We found 85 radio counterparts to 66 SCUBA-2 sources. There are nine SCUBA-2 sources with two radio counterparts and five with three radio counterparts. In the case of the main sample, 54 out of 83 SCUBA-2 sources matched with 70 radio sources and there are five/six triple/double-counterpart cases. For the supplementary sample, 12 out of 31 SCUBA-2 sources matched with 15 radio sources including three double counterparts. A catalog of the S2TDF submm sources is presented, including those that were matched with radio and optical/NIR sources.

We investigated the cumulative and differential SMG number counts as a function of the flux densities. Overall, the trends are similar to the results from other studies, but there is a slight excess at higher flux densities ($S_{850} > 10$ mJy), which arises from four bright sources in our relatively small survey area.

There are 51 (61) radio-located S2TDF SMGs with optical or NIR counterparts at S/N > 4 (S/N > 3.5). SED fitting was performed to these sources to derive their properties. We found that their estimated photometric redshifts have a median of $z = 2.22 \pm 0.12$ ($z = 1.96 \pm 0.08$), their median SFR is $300 \pm 40\,M_\odot\,\mathrm{yr}^{-1}$ ($240 \pm 40\,M_\odot\,\mathrm{yr}^{-1}$), and they have stellar masses of $1.7 \pm 0.3 \times 10^{11}\,M_\odot$ ($1.6 \pm 0.3 \times 10^{11}\,M_\odot$) for the sample at S/N > 4.0 (S/N > 3.5). This demonstrates that these galaxies lie above the main sequence at their redshifts, and that they represent the high-mass end of the star-forming population, suggesting that they are progenitors of massive early-type galaxies today.

We estimated dust masses and DSRs for our SMG sample, finding large dust masses, $M_d = 5.9 \pm 0.7 \times 10^8\,M_\odot$, which imply correspondingly high cold gas masses, $M_g \sim 2\text{--}3 \times 10^{10}\,M_\odot$, much higher than expected for typical field galaxies. These large reservoirs of gas are the fuel that drives the intense activity we see in the SMG population.

Our study provides a submm data set that is critical for investigating the obscured star formation activity of galaxies out to high redshifts within the JWST-TDF. Together with the anticipated JWST data, we expect that the S2TDF will provide valuable insights into the physical mechanisms that triggered vigorous star formation activity in the early universe, by combining the morphological information from JWST and the extreme galaxies identified by the S2TDF.

We thank Rick Perley and Ken Kellermann for their help with the VLA observations and their reduction and analysis. This work was supported by National Research Foundation of Korea (NRF) grants, No. 2020R1A2C3011091 and No. 2021M3F7A1084525, funded by the Korean government (MSIT). M.H. acknowledges support from a Korea Astronomy and Space Science Institute grant funded by the Korean government (MSIT) (No. 2022183005) and support from the Global PhD Fellowship Program through the NRF funded by the Ministry of Education (NRF-2013H1A2A1033110). I.R.S. acknowledges support from STFC (ST/T000244/1). R.A.W., S.H.C., and R.A.J. acknowledge support from NASA JWST Interdisciplinary Scientist grants NAG5-12460, NNX14AN10G, and 80NSSC18K0200 from GSFC. C.N.A.W. acknowledges funding from Hubble Space Telescope grant GO-15278 and the NIRCam Development Contract NAS5-02105 from NASA to the University of Arizona. H.S. acknowledges support from NRF grant No. 2021R1A2C4002725 and No. 2022R1A4A3031306. Y.M. acknowledges support from JSPS KAKENHI grant Nos. 17KK0098, 19H00697, 20H01953, 21H01133, 21H04489, and 22H01273. The James Clerk Maxwell Telescope is operated by the East Asian Observatory on behalf of The National Astronomical Observatory of Japan; Academia Sinica Institute of Astronomy and Astrophysics; the Korea Astronomy and Space Science Institute; Center for Astronomical Mega-Science (as well as the National Key R&D Program of China with No. 2017YFA0402700). Additional funding support is provided by the Science and Technology Facilities Council of the United Kingdom and participating universities in the United Kingdom and Canada. Additional funds for the construction of SCUBA-2 were provided by the Canada Foundation for Innovation. The National Radio Astronomy Observatory is a facility of the National Science Foundation operated under cooperative agreement by Associated Universities, Inc. Based in part on data collected at the Subaru Telescope and retrieved from the HSC data archive system, which is operated by the Subaru Telescope and Astronomy Data Center at the National Astronomical Observatory of Japan. Observations reported here were obtained at the MMT Observatory, a joint facility of the Smithsonian Institution and the University of Arizona.

*Facilities:* JCMT, VLA, Subaru, MMT.

# Appendix A
# VLA Observations of the JWST-TDF

## A.1. Introduction

The VLA observations presented here were obtained as part of an ongoing radio program using the VLA and the Very Long Baseline Array (VLBA) to generate a list of extragalactic radio sources in the JWST-TDF using the VLA, and then identify those containing a significant AGN component using the VLBA. The VLBA can observe multiple targets in the primary beam of the antennas; however, it is impractical to image the entire beam so accurate positions of the targets must be known in advance. To this end the VLA observations of the JWST-TDF region are centered on the very compact, 200 mJy quasar J1723+6547, for subsequent use as a phase reference for the VLBA observations (the chosen pointing contains no other strong radio sources).

However, these VLA observations also provide sensitive high-resolution radio coverage of the entire JWST-TDF as shown in Figure [1](#). This provides an excellent resource for studying the radio properties of other populations uncovered in this field; in particular these high-resolution observations can be used to identify likely counterparts to the submm sources detected in our low-resolution SCUBA-2 map of this field.

## A.2. Observations and Data Reduction

The observations were undertaken using the VLA "*S* band" ($\nu = 1.989\text{--}4.013$ GHz), which is optimal for this project as it provides a good compromise between high sensitivity and a wide field of view. Since the majority of galaxies observed have steep spectra, the relatively low frequency also helps detectability. The single pointing gives the lowest detection level for a given total exposure near the pointing center, but at the cost of the sensitivity dropping away from this field center.

We used the VLA to observe a field centered on J1723 +6547 (17 23 14.1381, +65 47 46.179) using the *S*-band receivers, with the 1.989–4.013 GHz frequency range divided

**Table 2**
Log of the VLA Observations of the JWST-TDF

| Date | Config. | Time$_{obs}$ (hr) | $S_{J1800}$ (Jy) | $\delta S_{J1800}$ (Jy) | $S_{J1723}$ (Jy) | $\delta S_{J1723}$ (Jy) |
|---|---|---|---|---|---|---|
| 2017 Nov 29 | B | 2 | 1.810 | 0.011 | 0.214 | 0.001 |
| 2017 Dec 18 | B | 2 | 1.680 | 0.013 | 0.194 | 0.001 |
| 2018 Jun 01 | A | 2 | 2.188 | 0.007 | 0.212 | 0.001 |
| 2018 Jun 02 | A | 4 | 2.144 | 0.007 | 0.219 | 0.001 |
| 2018 Jun 03 | A | 4 | 2.138 | 0.005 | 0.219 | 0.001 |
| 2018 Jun 04 | A | 4 | 2.136 | 0.002 | 0.222 | 0.001 |
| 2018 Jun 05 | A | 10 | 2.152 | 0.004 | 0.227 | 0.001 |
| 2018 Jun 06 | A | 4 | 2.178 | 0.002 | 0.228 | 0.001 |
| 2018 Jun 09 | A | 4 | 2.164 | 0.004 | 0.235 | 0.001 |
| 2018 Jun 10 | A | 6 | 2.138 | 0.004 | 0.231 | 0.001 |
| 2018 Jun 11 | A | 6 | 2.099 | 0.009 | 0.226 | 0.001 |

**Note.** We list (1) the date of the observations, (2) the VLA configuration employed, (3) the total observing time, (4–5) the calibration flux density adopted for J1800+7828 and its uncertainty, and (6–7) the flux density of J1723+6547 and its uncertainty.

into 16 contiguous subbands each of width $\Delta\nu = 128$ MHz. Each subband was further divided into 64, 2 MHz channels. Observations consisted of 2 hr blocks, one or more of which were observed on a given day. All data taken on a given calendar day were processed together. The observations are summarized in Table 2, which gives the date, VLA configuration, and total duration of the observations. Each 2 hr block included the astrometric calibrator J1800+7828 and one of the photometric/bandpass/polarization calibrators J1331+3030 (3C 286) and J0137+3309 (3C 48).

### A.2.1. Calibration

Calibration and imaging used the OBIT package (Cotton 2008).[15] Due to the strong source at the field center, special care was needed to minimize artifacts. Calibration and editing was done on each day's data independently and consisted of the steps described below. As the target field was dominated by a strong, compact source it was included as a calibrator. In each step deviant calibration solutions were detected and flagged along with the corresponding data. Standard structural and spectral models for J1331+3030 and J0137+3309 (Perley & Butler 2013a) were used as appropriate. Processing consisted of the following:

1. Fixed flagging: Frequency ranges known to contain strong, persistent radio frequency interference (RFI) were flagged.
2. Initial flagging: Running medians in time and frequency were used on the data to identify and flag RFI.
3. Initial amplitude calibration: Amplitude corrections were determined from the "switch power" calibration signals injected into each data stream.
4. Delay calibration: Residual group delays were determined for all calibrators and the target field.
5. Bandpass calibration: Amplitude and phase correction spectra were determined from the bandpass calibrator.
6. Amplitude and phase calibration: Complex gain solutions for the calibrators were determined for each calibrator and the target field. The daily flux density of the phase reference source was determined by a comparison of gain solutions with the photometric standard(s) and was used to correct the amplitudes of the solutions for it and the target. This value is given in Table 2.
7. Flagging of calibrated data: Flagging operations for which calibrated data were needed were done.
8. Repeat: The flaggings from the above steps were kept and the calibration was repeated.
9. Polarization calibration: Instrumental polarization was determined from the phase reference calibrator when an adequate range of parallactic angles was available; solutions were done in 16 MHz blocks. The cross-hand delay and phase were determined from J1331+3030 (preferred) or J0137+3309 (Perley & Butler 2013b).

All data sets were then subjected to a baseline-dependent time averaging to reduce the data volume. This averaging was subject to the constraint that the amplitudes were reduced by no more than 1% to a radius of 12′ and averaging was no more than 0.35 minutes.

### A.2.2. Imaging

As the observations were made over an extended period and the central source is mildly variable (see calibration values in Table 2), data from each day's observations were imaged and the contribution of the central source was subtracted. Initial imaging included a phase+delay self-calibration followed by an amplitude and phase self-calibration. The imaging used the OBIT task MFIMAGE, which is described in more detail in Cotton et al. (2018). Significant artifacts survived this process and additional filtering was needed to suppress them.

### A.2.3. Image Artifacts

The imaging artifacts arising from the strong central source included the following pathologies, given with their remediation (when available):

1. "Fingerprints"—Spiral patterns resembling fingerprints were traced to minor residual group delay errors when the group delay corrections were determined solely from the calibrators. Including the target in the group delay calibration and doing phase and delay self-calibration in the imaging largely eliminated these.
2. "Black stripes"—Initial imaging resulted in dark horizontal bands through and near the central source. These were traced to the periods of time when the fringe rate on a given baseline went through zero allowing greater sensitivity to RFI. As these are near $u = 0$ the resulting artifacts appear as horizontal stripes at the field center. These artifacts were largely suppressed by subtracting the image sky model from the data, averaging the data, and clipping the data above a given level.
3. "Bowls"—All of the "A" configuration data sets but neither of the "B" data sets show a negative bowl around the position of the central source with a maximum depth of $\sim 150\,\mu$Jy, even after the source was subtracted. The source of this artifact has not been determined but it only extends a few beams from the central source. As the central source is completely unresolved, this bowl is not the feature commonly seen around very extended, bright features.

---
[15] http://www.cv.nrao.edu/~bcotton/Obit.html

Table 3
VLA 3 GHz Source Catalog

| ID | R.A. (J2000) | δR.A. (″) | Decl. (J2000) | δDecl. (″) | $S_{3GHz}$ (μJy) | $δS_{3GHz}$ (μJy) | Size (″) | δSize (″) | Minor (″) | δMinor (″) | P.A. (deg) | δP.A. (deg) |
|---|---|---|---|---|---|---|---|---|---|---|---|---|
| 1 | 17 21 18.6268 | 0.0084 | +65 47 38.086 | 0.055 | 63.7 | 8.0 | <0.66 | ⋯ | ⋯ | ⋯ | ⋯ | ⋯ |
| 2 | 17 21 20.0334 | 0.0129 | +65 50 54.965 | 0.095 | 45.8 | 9.1 | <0.90 | ⋯ | ⋯ | ⋯ | ⋯ | ⋯ |
| 3 | 17 21 24.2805 | 0.0170 | +65 48 50.988 | 0.082 | 88.9 | 15.6 | 1.00 | 0.24 | <0.79 | ⋯ | −57 | 12 |
| 4 | 17 21 26.4499 | 0.0216 | +65 47 57.197 | 0.079 | 1521.6 | 15.1 | LAS 16.2 | ⋯ | ⋯ | ⋯ | ⋯ | ⋯ |
| 5 | 17 21 28.3552 | 0.0151 | +65 46 59.518 | 0.075 | 36.2 | 6.1 | <1.10 | ⋯ | ⋯ | ⋯ | ⋯ | ⋯ |

**Note.** The central source, J1723+6547, was added to the catalog in spite of being removed from the image. We list (1) the source ID; (2–5) the J2000 positions of each source and their uncertainties; (6–7) the 3 GHz flux density, corrected for the primary beam, and its uncertainty; and (8–13) the deconvolved angular size and/or upper limit or largest angular size (LAS) and the position angle.

(This table is available in its entirety in machine-readable form.)

### A.2.4. Combined Image

After each individual day's data sets were imaged, the core was subtracted, and the artifact filtering was run, the data were combined into a single set and imaged with no further self-calibration. CLEANing proceeded to a minimum residual of 5 μJy beam$^{-1}$ or a maximum of 20,000 components. Imaging used a robust factor of 0.0 and the resulting images used a circular Gaussian restoring beam of a 0″.7 FWHM. The rms in quiet places in the resultant image is 0.95 μJy beam$^{-1}$ and the brightest pixel is 3.1 mJy beam$^{-1}$.

### A.2.5. Source Catalog

The construction of the source list used the OBIT task FNDSOU, which fits elliptical Gaussians to components identified in islands of emission. The derived list was compared with the image and entries corresponding to clear artifacts or multiple entries from an extended source were removed. Error analysis followed the development of Condon (1997). The field image FITS file and the full source list for 756 sources with a peak brightness >5 times the local rms and with a primary-beam gain of at least 10% at 3 GHz (within a radius of 12′) are given online with a sample shown in Table 3.

Some sources are sufficiently extended to not be modeled by a simple Gaussian. The flux densities of these were determined by an integral of the image pixels bounded by a rectangular box. The LAS of the source was derived from the diagonal extent of this rectangle. These sources are indicated in Table 3 by the LAS given in the "Size" column.

### A.3. Summary

Our VLA 3 GHz observations of the JWST-TDF reach an rms of 1 μJy beam$^{-1}$ in the region around the bright radio source J1723+6547 at a resolution of 0″.7. The catalog of sources above 5σ generated from this map contains 756 sources, which we used to identify the counterparts to the SCUBA-2 sources in this region. The full catalog of 3 GHz sources will also be employed in our ongoing VLBA observations of sources in this field.

## Appendix B
## Jackknife Simulation and Flux Distribution

As we explained in Section 3.2, we performed 1000 jackknife simulations to determine the amount of flux boosting and the completeness of the S2TDF. Here, we present two-dimensional density plots showing the simulation results (Figure 15).

Figure 16 presents the flux density distribution of the observed flux and the deboosted flux from the empirical recovery method.

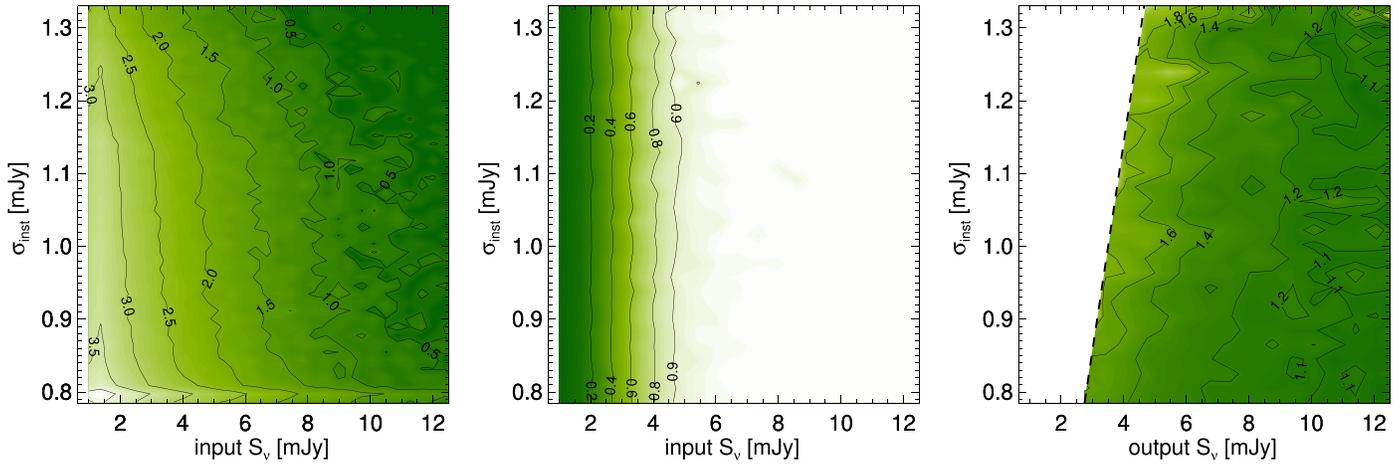

**Figure 15.** Two-dimensional density plots showing the results of 1000 jackknife simulations performed for our survey. (Left) A plot of injected sources as a function of input flux density and instrumental noise. (Middle) A density plot of completeness, the ratio of the number of recovered sources to that of injected sources, as a function of input flux density and instrumental noise. The contours are labeled with the completeness fraction. (Right) The plot shows the average boosting factor of the output flux density with the given instrumental noise. The dashed line shows the S/N = 3.5 limit. The contour labels indicate the flux boost factor.

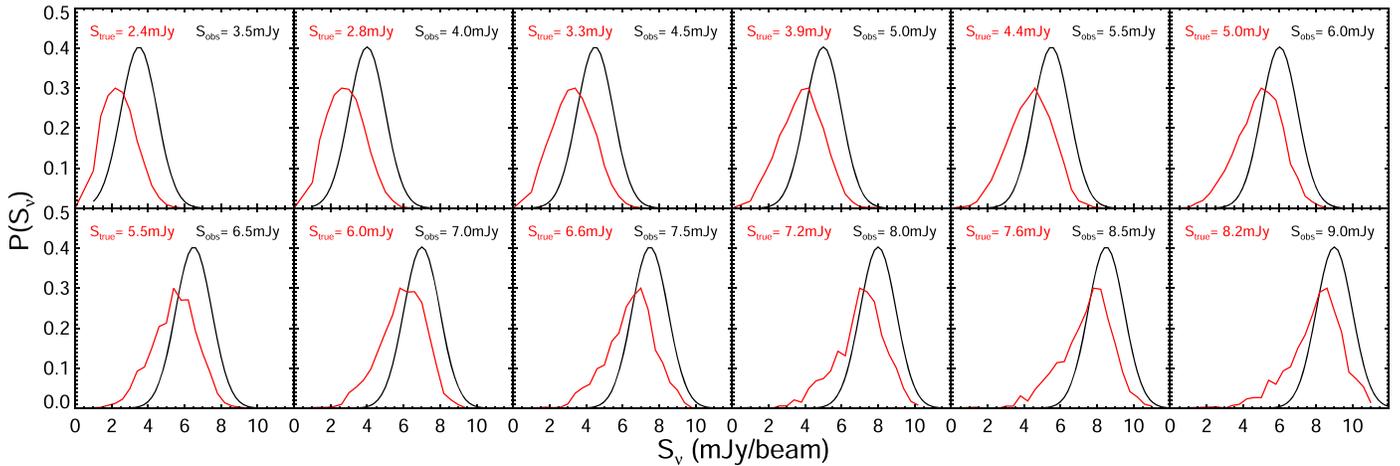

**Figure 16.** Flux density distribution of observed flux (black curve) and deboosted flux (red curve) for example sources derived from the empirical recovery method. Labels with red and black colors show the median value of the corrected flux and observed flux, respectively. All results assume Gaussian uncertainties of $\sigma_{inst} = 1.0$ mJy, which is the median instrumental noise of the final map of the TDF. For bright sources, the effect of flux boosting is minimal. But, at the limit of our catalog, it can be considerable ($\simeq 30\% - 50\%$).

## Appendix C
## Optical Spectroscopy

The spectroscopic data were obtained using Binospec (Fabricant et al. 2019) at the MMT Observatory. These were obtained using the 270 line mm$^{-1}$ grating, which allows coverage of approximately 4000–9000 Å with a typical dispersion of 1.30 Å pixel$^{-1}$ and a resolution of 1300. The sample of galaxies was constructed by combining a preliminary catalog from the HEROES Subaru HSC imaging (G. Hasinger, private communication) and a catalog derived from MMT/MMIRS NIR imaging (C. N. A. Willmer et al. 2022, in preparation), limited at $r \sim 23$. The slit assignment was prioritized for sources with X-ray, VLA data, or submm counterparts in preliminary catalogs coming from members of the collaboration (private communication from W. P. Maksym, R. Windhorst, or I. R. Smail, respectively).

The data were reduced using a specially designed pipeline that produces wavelength- and flux-calibrated one-dimensional spectra (Kansky et al. 2019). Redshifts were measured using custom-written code using a combination of real-space cross-correlation and emission line fits. All spectra were visually inspected (by C. N. A. Willmer) and a redshift quality was assigned using the same criteria adopted by the DEEP2 survey (Newman et al. 2013), where a redshift quality of 4 is likely to be correct at a 95% level or better and quality 3 is correct at a level of ~90%. Data qualities less than 3 were not considered in the analyses. The output from each of the observed fields was then merged into a single list using the positions of the NIR imaging, which are tied to the Gaia Data Release 2 reference frame.

## Appendix D
## The Best-fit MAGPHYS Model SEDs for the S2TDF SMGs

Figure 17 displays the best-fit MAGPHYS model SEDs for the 85 SMGs.

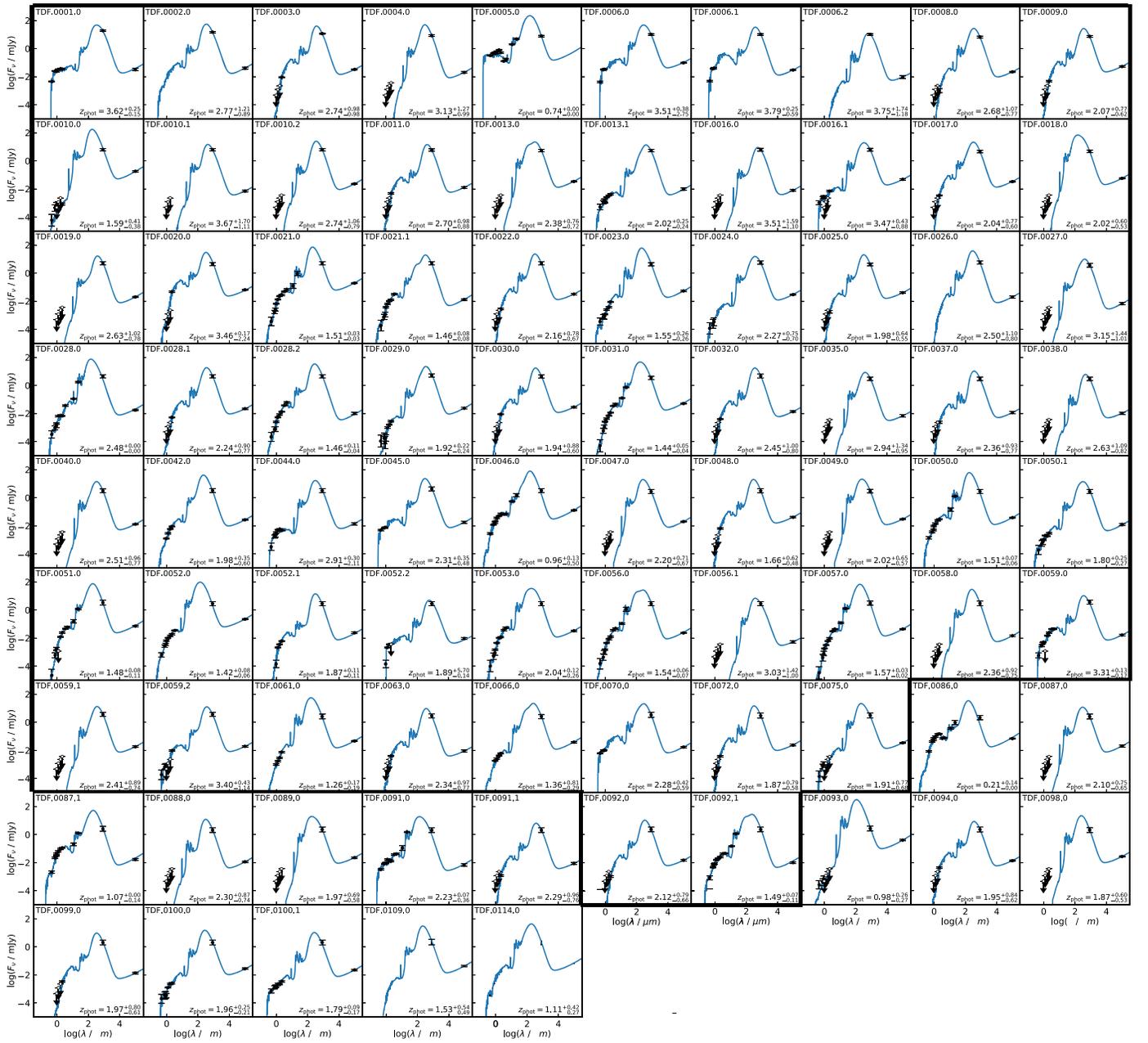

**Figure 17.** The best-fit MAGPHYS model SEDs for the 85 SMGs in this work that have 3 GHz radio counterparts, including the 61 sources for which we have sufficient optical and/or NIR photometry data to derive useful constraints on their SED properties. The bold frame represents the main targets at S/N > 4. We list the desired photometric redshifts and if available any spectroscopic values. The sources range within $z = 0.8\text{–}3.8$ and all sources show a strong FIR peak in their SEDs traced by the SCUBA-2 850 $\mu$m emission.

## Appendix E
## S2TDF Source Catalog and Stamp Images

Figure 18 provides, in an online-only format, postage-stamp images for all the main and supplementary SCUBA-2 sources detected in the JWST-TDF. The first 10 online-only figures are of the main set; the last four online-only figures are of the supplement. Table 4 provides a detailed description of the full version of JWST-TDF SCUBA-2 850 $\mu$m. The catalog is provided in FITS format and includes both the main and supplementary sources, which are differentiated by the MAIN Boolean column.

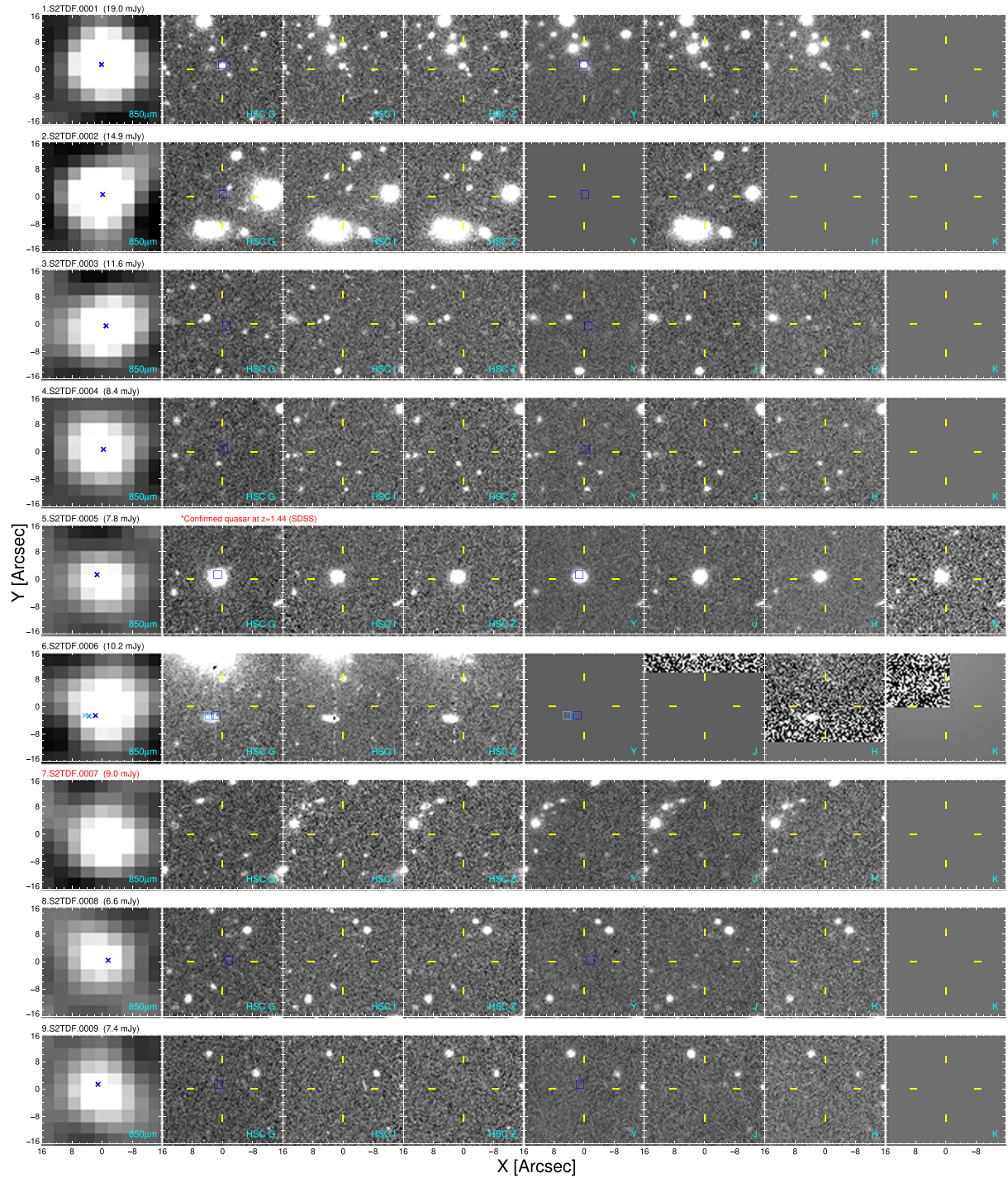

**Figure 18.** The $32'' \times 32''$ images of the main 83 SCUBA-2 sources (S/N > 4) detected in the JWST-TDF. If the source has a radio counterpart, the label is black; otherwise, it is red. The positions of the radio counterparts are shown with blue crosses on the JCMT image and square marks on the HSC $G$ and MMIRS $Y$ and $K$ images. If there is more than one counterpart lighter blue marks are also added to the figure in orders of distances. The matching radius of $7''\!.5$ is indicated with four yellow marks. The complete set of postage stamps (14 images) is available online. The first 10 images are of the main sources; the last four are of the supplementary sources.

(An extended version of this figure is available.)

Table 4
Description of the Full Version of JWST-TDF SCUBA-2 850 μm Source Catalogs

| Index | Label | Units | Description |
|---|---|---|---|
| 1 | index | ⋯ | Index |
| 2 | JCMT_index | ⋯ | JCMT index |
| 3 | JCMT_ID | ⋯ | IAU position–based name, S2TDFJHHMMSS+/−DDMMSS |
| 4 | JCMT_SHORT_ID | ⋯ | Short name, S2TDF.JCMT_index |
| 5 | ID | ⋯ | Identifier with VLA multiplicity index, JCMT_SHORT_ID.[0, 1, 2] |
| 6 | MAIN | Boolean | True = main (JCMT_S/N > 4); False = supplementary (JCMT_S/N = 3.5–4.0) |
| 7 | Got_VLA_ID | Boolean | True = VLA match |
| 8 | NOT_Got_VLA_ID | Boolean | True = no VLA match |
| 9 | Closest_ID | Boolean | True = closest matched object |
| 10 | Multiple_IDs | Boolean | True = multiple VLA matches |
| 11 | N(S>S) | ⋯ | Radio source number density $S_{3GHz} > 10$ μJy |
| 12 | Ni | ⋯ | Radio source number density $S_{3GHz} > S_i$ |
| 13 | Pc | ⋯ | $P_c = \pi r_s^2 N_T$, Equation (8) |
| 14 | P* | ⋯ | $P^* = \pi r_i^2 N_i$, Equation (9) |
| 15 | p | ⋯ | Corrected Poissonian probability, Equation (6) |
| 16 | n_radio_counterpart | ⋯ | Number of radio counterparts |
| 17 | closest_rank | ⋯ | Rank of the radio-matched object based on the separated distance [1, 2, 3] |
| 18 | JCMT_FLUX_PEAK | mJy | Peak JCMT 850 μm flux |
| 19 | JCMT_RMS_PEAK | mJy | rms uncertainty in peak flux |
| 20 | JCMT_FLUX_MODEL | mJy | Model JCMT 850 μm flux |
| 21 | JCMT_RMS_MODEL | mJy | rms uncertainty in model flux |
| 22 | JCMT_S/N | ⋯ | S/N, JCMT 850 μm |
| 23 | JCMT_BLEND | ⋯ | [0, 1], 1 = blend |
| 24 | JCMT_FLUX_DEBOOST | mJy | Deboosted JCMT 850 μm flux |
| 25 | JCMT_FLUX_DEBOOST_ERRLO | mJy | Lower uncertainty in JCMT_FLUX_DEBOOST |
| 26 | JCMT_FLUX_DEBOOST_ERRHI | mJy | Upper uncertainty in JCMT_FLUX_DEBOOST |
| 27 | JCMT_TOTAL_ERRLO | mJy | Lower total uncertainty in JCMT_FLUX_DEBOOST |
| 28 | JCMT_TOTAL_ERRHI | mJy | Upper total uncertainty in JCMT_FLUX_DEBOOST |
| 29 | JCMT_RA_PEAK_DEG_FIN | deg | R.A., peak flux, decimal degrees |
| 30 | JCMT_DEC_PEAK_DEG_FIN | deg | Decl., peak flux, decimal degrees |
| 31 | JCMT_RA_MODEL_DEG_FIN | deg | R.A., peak model flux, decimal degrees |
| 32 | JCMT_DEC_MODEL_DEG_FIN | deg | Decl., peak model flux, decimal degrees |
| 33 | ID_VLA | ⋯ | Identifier from VLA catalog (Appendix A) |
| 34 | RA_VLA | deg | R.A., VLA source, decimal degrees |
| 35 | DEC_VLA | deg | Decl., VLA source, decimal degrees |
| 36 | radio_flux | μJy | 3 GHz radio flux, $S_{3GHz}$ |
| 37 | dS3GHz | μJy | Uncertainty in 3 GHz flux |
| 38 | angular_distance | arcsec | Angular separation between VLA and JCMT |
| 39 | sed_group | ⋯ | 1: K detection + opt./NIR data; 2: no K detection; 3: VLA+submm data only |
| 40 | GMAG_APER_300 | mag | G-band aperture magnitude from HSC (Bosch et al. 2018; Miyazaki et al. 2018) |
| 41 | GMAGERR_APER_300 | mag | Uncertainty in GMAG_APER_300 |
| 42 | IMAG_APER_300 | mag | I-band aperture magnitude from HSC (Bosch et al. 2018; Miyazaki et al. 2018) |
| 43 | IMAGERR_APER_300 | mag | Uncertainty in IMAG_APER_300 |
| 44 | ZMAG_APER_300 | mag | Z-band aperture magnitude from HSC (Bosch et al. 2018; Miyazaki et al. 2018) |
| 45 | ZMAGERR_APER_300 | mag | Uncertainty in ZMAG_APER_300 |
| 46 | YMAG_APER_3 | mag | Y-band aperture magnitude from MMIRS (McLeod et al. 2012) |
| 47 | YMAGERR_APER_3 | mag | Uncertainty in YMAG_APER_3 |
| 48 | JMAG_APER_3 | mag | J-band aperture magnitude from MMIRS (McLeod et al. 2012) |
| 49 | JMAGERR_APER_3 | mag | Uncertainty in JMAG_APER_3 |
| 50 | HMAG_APER_3 | mag | H-band aperture magnitude from MMIRS (McLeod et al. 2012) |
| 51 | HMAGERR_APER_3 | mag | Uncertainty in HMAG_APER_3 |
| 52 | KMAG_APER_3 | mag | K-band aperture magnitude from MMIRS (McLeod et al. 2012) |
| 53 | KMAGERR_APER_3 | mag | Uncertainty in KMAG_APER_3 |
| 54 | w1_ab | mag | WISE band 1 AB magnitude (Wright et al. 2010) |
| 55 | dw1 | mag | Uncertainty in w1_ab |
| 56 | w2_ab | mag | WISE band 2 AB magnitude (Wright et al. 2010) |
| 57 | dw2 | mag | Uncertainty in w2_ab |
| 58 | w3_ab | mag | WISE band 3 AB magnitude (Wright et al. 2010) |
| 59 | w3msigmpro | mag | Uncertainty in w3_ab |
| 60 | w4_ab | mag | WISE band 4 AB magnitude (Wright et al. 2010) |
| 61 | w4msigmpro | mag | Uncertainty in w4_ab |
| 62 | Major | arcsec | Ellipse-fit major axis, arcseconds |
| 63 | dMajor | arcsec | Uncertainty in Major: 99 = LAS; −99 = upper limit |

**Table 4**
(Continued)

| Index | Label | Units | Description |
|---|---|---|---|
| 64 | Minor | arcsec | Ellipse-fit minor axis, arcseconds |
| 65 | dMinor | arcsec | Uncertainty in Minor: −99 = upper limit |
| 66 | PA | deg | Ellipse-fit position angle |
| 67 | dPA | deg | Uncertainty in PA |
| 68 | chi | ⋯ | $\chi^2$ of the SED fitting result |
| 69 | Mstar_l | $\log(M_\odot)$ | Lower bound on Mstar |
| 70 | Mstar | $\log(M_\odot)$ | Stellar mass in log |
| 71 | Mstar_u | $\log(M_\odot)$ | Upper bound on Mstar |
| 72 | L_dust_l | $\log(L_\odot)$ | Lower bound on L_dust |
| 73 | L_dust | $\log(L_\odot)$ | Dust luminosity, $L_{\rm dust}$, in log |
| 74 | L_dust_u | $\log(L_\odot)$ | Upper bound on L_dust |
| 75 | Tdust_l | K | Lower bound on Tdust |
| 76 | Tdust | K | Kinetic dust temperature, $T_{\rm dust}$ |
| 77 | Tdust_u | K | Upper bound on Tdust |
| 78 | SFR_l | $\log(M_\odot\,{\rm yr}^{-1})$ | Lower bound on SFR |
| 79 | SFR | $\log(M_\odot\,{\rm yr}^{-1})$ | Star formation rate in log |
| 80 | SFR_u | $\log(M_\odot\,{\rm yr}^{-1})$ | Upper bound on SFR |
| 81 | Mdust_l | $\log(M_\odot)$ | Lower bound on Mdust |
| 82 | Mdust | $\log(M_\odot)$ | Dust mass in log |
| 83 | Mdust_u | $\log(M_\odot)$ | Upper bound on Mdust |
| 84 | Age_l | log(yr) | Lower bound on Age |
| 85 | Age | log(yr) | Age in log |
| 86 | Age_u | log(yr) | Upper bound on Age |
| 87 | AV_l | mag | Lower bound on AV |
| 88 | AV | mag | $V$-band reddening |
| 89 | AV_u | mag | Upper bound on AV |
| 90 | tau_l | Gyr | Lower bound on tau |
| 91 | tau | Gyr | Star formation timescale parameter |
| 92 | tau_u | Gyr | Upper bound on tau |
| 93 | sSFR_l | $\log({\rm yr}^{-1})$ | Lower bound on sSFR |
| 94 | sSFR | $\log({\rm yr}^{-1})$ | Specific SFR in log |
| 95 | sSFR_u | $\log({\rm yr}^{-1})$ | Upper bound on sSFR |
| 96 | z_lower | ⋯ | Lower bound on z_med |
| 97 | z_med | ⋯ | Median photometric redshift |
| 98 | z_upper | ⋯ | Upper bound on z_med |
| 99 | eb_l | mag | Lower bound on eb |
| 100 | eb | mag | Color excess, $E(B-V)$ |
| 101 | eb_u | mag | Upper bound on eb |
| 102 | Notes | ⋯ | Notes |

**Note.** The full version of the JWST-TDF SCUBA-2 catalog for the main (S/N > 4) sources includes VLA counterparts from the Poisson probability crossmatching (see Section 4.1.1). The calculated Poissonian probabilities are presented as $p$. If there are multiple counterparts, we list them in the order of probability. The number after the decimal point in "ID" indicates multiple radio counterparts to a common submm source. The full catalog, including both the main and supplementary (S/N = 3.5–4.0) sources, is available online in FITS format.

(This table is available in its entirety in machine-readable form.)

## ORCID iDs

Minhee Hyun ● https://orcid.org/0000-0003-4738-4251
Myungshin Im ● https://orcid.org/0000-0002-8537-6714
Ian R. Smail ● https://orcid.org/0000-0003-3037-257X
William D. Cotton ● https://orcid.org/0000-0001-7363-6489
Jack E. Birkin ● https://orcid.org/0000-0002-3272-7568
Satoshi Kikuta ● https://orcid.org/0000-0003-3214-9128
Hyunjin Shim ● https://orcid.org/0000-0002-4179-2628
Christopher N. A. Willmer ● https://orcid.org/0000-0001-9262-9997
James J. Condon ● https://orcid.org/0000-0003-4724-1939
Rogier A. Windhorst ● https://orcid.org/0000-0001-8156-6281
Seth H. Cohen ● https://orcid.org/0000-0003-3329-1337
Rolf A. Jansen ● https://orcid.org/0000-0003-1268-5230
Chun Ly ● https://orcid.org/0000-0002-4245-2318
Yuichi Matsuda ● https://orcid.org/0000-0003-1747-2891
Giovanni G. Fazio ● https://orcid.org/0000-0002-0670-0708
A. M. Swinbank ● https://orcid.org/0000-0003-1192-5837
Haojing Yan ● https://orcid.org/0000-0001-7592-7714